\documentclass[11pt]{article}
\usepackage{graphicx,amsmath,amsfonts,amssymb,bm,hyperref,url,breakurl,epsfig,epsf,color,fullpage,MnSymbol,mathbbol, fmtcount,algorithmic,algorithm, semtrans
} 

\usepackage{titlesec}
\usepackage{abstract}

\usepackage{tikz}
\usepackage{pgfplots}
\usepgfplotslibrary{groupplots}
\usetikzlibrary{positioning}
\usepackage{subcaption}

\usetikzlibrary{pgfplots.groupplots}

\titleformat{\paragraph}
{\normalfont\normalsize\bfseries}{\theparagraph}{1em}{}
\titlespacing*{\paragraph}
{0pt}{3.25ex plus 1ex minus .2ex}{1.5ex plus .2ex}

\usepackage{caption,subcaption}

\usepackage[bottom,hang,flushmargin]{footmisc} 

\usepackage{hyperref}
\definecolor{darkred}{RGB}{150,0,0}
\definecolor{darkgreen}{RGB}{0,150,0}
\definecolor{darkblue}{RGB}{0,0,200}
\hypersetup{colorlinks=true, linkcolor=darkred, citecolor=darkgreen, urlcolor=black}

\newtheorem{theorem}{Theorem}
\newtheorem{lemma}{Lemma}

\newtheorem{definition}{Definition}

\newtheorem{remarks}{Remarks}

\newcommand{\fronorm}[1]{\left\|#1\right\|_{F}}
\newcommand{\onenorm}[1]{\left\|#1\right\|_{\ell_1}}
\newcommand{\twonorm}[1]{\left\|#1\right\|_{\ell_2}}

\newcommand{\abs}[1]{\left|#1\right|}

\newcommand{\R}{\mathbb{R}}

\newcommand{\E}{\operatorname{\mathbb{E}}}

\newcommand{\vct}[1]{\bm{#1}}
\newcommand{\mtx}[1]{\bm{#1}}
\definecolor{ejc}{RGB}{0,0,255}

\numberwithin{equation}{section} 

\def \endprf{\hfill {\vrule height6pt width6pt depth0pt}\medskip}

\renewcommand{\algorithmicrequire}{\textbf{Input:}}
\renewcommand{\algorithmicensure}{\textbf{Output:}}

\begin{document}

\title{Fundamental Resource Trade-offs for Encoded Distributed Optimization}

\author{A.~Salman~Avestimehr, Seyed~Mohammadreza~Mousavi~Kalan, and Mahdi~Soltanolkotabi\\ 
Ming Hsieh Department of Electrical Engineering, University of Southern California \\Email: avestimehr@ee.usc.edu, mmousavi@usc.edu, soltanol@usc.edu}

\date{\vspace{-5ex}}
\maketitle
\begin{abstract}
Dealing with the shear size and complexity of today's massive data sets requires computational platforms that can analyze data in a parallelized and distributed fashion. A major bottleneck that arises in such modern distributed computing environments is that some of the worker nodes may run slow. These nodes a.k.a.~stragglers can significantly slow down computation as the slowest node may dictate the overall computational time. A recent computational framework, called encoded optimization, creates redundancy in the data to mitigate the effect of stragglers. In this paper we develop novel mathematical understanding for this framework demonstrating its effectiveness in much broader settings than was previously understood. We also analyze the convergence behavior of iterative encoded optimization algorithms, allowing us to characterize fundamental trade-offs between convergence rate, size of data set, accuracy, computational load (or data redundancy), and straggler toleration in this framework.
\end{abstract}

\section{Introduction}\label{sec:intro}

Modern data sets are massive in size and complexity consisting of tens of billions of examples. These data sets are also very high-dimensional with numerous detailed information gathered for each example. Furthermore, due to the proliferation of a variety of personal devices many modern data sets are stored or collected in a distributed manner. To process such data sets in a timely manner, distributed computing algorithms/platforms that can analyze data in a parallelized or fully decentralized fashion are crucial. 

%Implementing inferential learning algorithms in modern distributed computing platform, however, pose new challenges. 

%One major challenge that arises when implementing learning algorithms in modern distributed computing environments, such as Amazon EC2, is that some of the nodes may run slow.
%These slow nodes a.k.a. stragglers are 

As we scale out computations across many distributed nodes in modern distributed computing environments, such as Amazon EC2, a major performance bottleneck is the latency in waiting for slowest nodes, or  ``stragglers'' to finish their tasks~\cite{dean2013tail}. These stragglers are caused by various forms of ``system noise'' (e.g., deallocation of computational resources, bandwidth limitation, node failure, etc) and can significantly slow down computation as the slowest node may dictate the overall computational time. The conventional approaches to mitigate the impact of stragglers involve creation of some form of ``computational redundancy''.  For example, \emph{replicating} the straggling task on another available node is a common approach to deal with stragglers (e.g.,~\cite{zaharia2008improving}).

%Stragglers are an Achilles heel of most distributed computing platforms. Indeed, as noted by \cite{}\MS{add a citation} stragglers can increase the run-time by a factor of ...\MS{Complete this}. One simple but common approach to deal with stragglers is to simply replicate tasks across multiple nodes \cite{Wang:2015}. This allows the computational tasks to be completed without having to wait for the straggler nodes. 

More recent approaches \cite{Lee:2016,li2016fundamental,li2016unified,NIPS2017_7027,dutta2016short,Tandon,Halbawi,Raviv} bring to bear ideas from coding theory to distributed computing. These \emph{coded computing} approaches create redundancy in the \emph{computation tasks} in unorthodox coded forms (as opposed to conventional replication approaches), thereby alleviating the effect of stragglers more efficiently. 

 Coded computing has also been proposed for creating redundancy in distributed optimization problems~\cite{Karakus,Karakus-NIPS}. The key idea of this approach, named \emph{encoded optimization}, is to linearly encode the data variables in the optimization.  The encoded data is then distributed across the computational nodes and distributed optimization algorithms are then applied to these encoded data. Due to the redundancy created in the data the optimization algorithms can be completed without having to wait for the straggler nodes.

%A more recent approach called encoded optimization \cite{Karakus,Karakus-NIPS} aims to create redundancy in the data set rather than the computational tasks. In this approach the data set is encoded prior to data distribution. The encoded data is then distributed across the computational nodes and distributed optimization algorithms are then applied to these encoded data. Due to the redundancy created in the data the optimization algorithms can be completed without having to wait for the straggler nodes.

The encoded optimization framework provides an intriguing approach to deal with the effect of stragglers. However, our mathematical understanding of the effect of this data encoding strategy is limited. In particular existing results such as \cite{Karakus,Karakus-NIPS} mostly focus on understanding the effect of random encoding strategies on the optimal solution to unconstrained least-squares problems. Furthermore, there is very limited understanding of how such encoding strategies affect the use of various computational and data resources. This is particularly important as in many modern applications ranging from imaging to online advertisement and financial trading we are interested in algorithms that can operate under multiple constraints (e.g. under a limited time budget). Efficient learning from encoded data under these constraints poses new challenges: How can we incorporate domain-specific prior knowledge in a principled manner? What algorithms should we use under a fixed time budget? How much of the data should we use? Should we use all of the data or just parts of it? How many passes (or iterations)
of the algorithm is required to get to an accurate solution? How much redundancy should we create in our data? How does the amount of redundancy present in our data encoding strategy affect the convergence behavior and run-time of our algorithms? How many straggler nodes can a particular form of data encoding approach tolerate?

At the heart of answering these questions is the ability to predict run-time of encoded optimization algorithms
as a function of the required accuracy, the size of data, the number of straggler nodes, the amount of prior knowledge, etc. That is, we need to understand precise trade-offs between run time, data size, accuracy, data redundancy and straggler toleration of iterative encoded optimization algorithms. In this paper we wish to precisely characterize such trade-offs, significantly broadening our current understanding of the encoded optimization paradigm. 
Our main contributions in this paper are as follows.

\begin{itemize}
    \item We study the encoded optimization framework in a much broader setting than previously understood. In particular, we demonstrate how prior knowledge can be incorporated in this framework via constraints on the optimization variables. Our guarantees are very general and can deal with arbitrary and potentially nonconvex constraints.
   
    \item Our results require a near minimal amount of data redundancy/replication (a.k.a.~computational load). We show that encoded optimization is effective as long as the data redundancy/replication exceeds (up to constants) the sum of the total number of stragglers and a precise quantity capturing the amount of prior knowledge that is enforced in the optimization algorithm. In fact, in certain cases our framework applies even when the number of encoded data is less than the number of data points allowing for data compression in lieu of redundancy/replication.

    \item We also precisely characterize the convergence rate of iterative encoded optimization algorithms as a function of various parameters including the straggler toleration, computational load, prior knowledge, as well as the size of the data set. This allows us to precisely characterize the various trade-offs between these fundamental resources.
\end{itemize}

\section{Problem formulation}

In this section we discuss the encoded distributed optimization framework and formulate the underlying fundamental resource trade-offs that we study in this paper.

%class of optimization problems/algorithms we are interested in solving and the various challenges and questions that arise in distributed implementations of these approaches. Section \ref{Psetting} introduces the class of optimization problems which are the focus of this paper along with a popular distributed implementation. We also discuss various challenges that arise when implementing these distributed approaches in modern computational platforms. Section \ref{EncOpt} introduces the encoded optimization framework aimed at addressing some of these challenges. Finally, Section \ref{trade} discusses the various quastions and resource trade-offs that arise in the encoded optimization framework.

% which we wish to characterize in this paper.

%distributed implementation distributed framework and the various challenges and question that arise when In this section we introduce the class of optimization problems we are interested in solving and how we can use distributed computing setup to solve these kinds of problems. But as will be explained, there are some challenges in exploiting distributed computing. We then introduce the idea of encoded optimization to combat the challenges arise in using distributed setups. Finally, we characterize the questions which using encoded optimization opens up, and explain the main purpose of this paper as well.

\subsection{Setting}

\label{Psetting}
In many modern applications in signal processing and machine learning we aim to infer models that best explain the training data. Given training data consisting of $n$ pairs of input features $\vct{x}_i\in\R^d$ and desired outputs $y_i\in\R$ we wish to infer a function that best explains the training data.The simplest functions are linear ones where the outputs are linear functions of the features. Specifically, we are interested in finding a parameter $\vct{\theta}^*\in\mathbb{R}^{d}$ obeying the following equations
\begin{align}
y_i=<\vct{x}_i,\vct{\theta}^*>+\vct{w}_i \quad\text{for}\quad i=1,2,...,n.
\end{align}
Here, $w_i$ denotes the noise present in our training examples. A natural approach to finding the best linear model is to minimize the empirical risk $\sum_{k=1}^n \ell(\langle\vct{x}_k,\vct{\theta}\rangle,y_k)$ via a quadratic loss $\ell(u,v)=\frac{1}{2}(u-v)^2$. This leads to the following optimization problem
\begin{align}
\label{original}
\begin{split}
&\hat{\vct{\theta}}=\underset{\vct{\theta}}{\arg\min}\text{ }\mathcal{L}(\vct{\theta}):=\frac{1}{2}\twonorm{\vct{y}-\mtx{X}\vct{\theta}}^2\\& \text{subject to}\quad \mathcal{R}(\vct{\theta})\le R.
\end{split}
\end{align}
Here, $\vct{y}\in\R^n$ is the output vector consisting of the outputs $\vct{y}=[y_1\ y_2\ ... \ y_n]^T$ and  $\mtx{X}\in\R^{n\times d}$ is the feature matrix consisting of the data features $\mtx{X}=[\vct{x}_{1}^{T}\ \vct{x}_{2}^{T}\ ...\ \vct{x}_{n}^{T}]^T$. Also, $\mathcal{R}:\R^d \rightarrow \R$ is a regularizer function that is used to avoid over-fitting and captures some notion of structure/complexity of the unknown parameter (with $R$ a tuning parameter). We note that while we will focus on linear models and quadratic losses many of the algorithms and technical proofs in this paper generalize to other models/losses. We aim to pursue these extensions in future publications.% publications. 

%fit is well-known approach to finding $$ 
%This problem is famous as linear regression which often consists of finding a vector $\hat{\vct{\theta}}\in\mathbb{R}^{d}$ that minimizes the least square error
%\begin{align}
%\hat{\vct{\theta}}=\underset{\vct{\theta}}{\arg\min}\sum_{i=1}^{n}(y_i-<\vct{x_i},\vct{\theta}>)^2
%\end{align}
%along with some constraints on the structure of the desired parameter.
%If we denote the output vector $\vct{y}=[y_1\ y_2\ ... \ y_n]^T$ and the feature matrix $\mtx{X}=[\vct{x}_{1}^{T}\ \vct{x}_{2}^{T}\ ...\ \vct{x}_{n}^{T}]^T$, we can reformulate the above optimization problem as follows

To solve optimization problems of the form \eqref{original} involving massive data sizes we need to utilize modern distributed computing platforms. While there are many popular distributed computing schemes \cite{Qi2017}, in this paper we focus on a OneToAll scheme which consists of a master node and $L$ workers. We focus on a distributed implementation of Projected Gradient Descent (PGD) for solving problems of the form \eqref{original}. To distribute PGD we assume the master partitions the features matrix $\mtx{}$ and the response vector $\vct{y}$ across rows between $L$ worker nodes. Specifically, we partition $\mtx{X}/\vct{y}$ into $L$ parts across rows with $\mtx{X}=\begin{bmatrix}\mtx{X}_1^T & \mtx{X}_2^T & \ldots & \mtx{X}_L^T\end{bmatrix}^T$, $\vct{y}=\begin{bmatrix}\vct{y}_1^T & \vct{y}_2^T & \ldots & \vct{y}_L^T\end{bmatrix}^T$. Here, $\mtx{X}_\ell\in\R^{n_\ell\times d}$ and $\vct{y}_\ell\in\R^{n_\ell}$ with $\sum_\ell n_\ell=n$. With this partition, worker $\ell$ receives data $\mtx{X}_\ell/\vct{y}_\ell$ and carries out updates/computations on this data. Starting from some initial solution $\vct{\theta}_0\in\R^d$, in each iterations the master sends the current update $\vct{\theta}_\tau$ to the workers. Each of the workers then calculates a partial gradient based on the portion of the data it has access to. Specifically, worker $\ell$ returns to the master the following partial gradient
\begin{align*}
\nabla \mathcal{L}^{(\ell)}(\vct{\theta})=\mtx{X}^T_{\ell}\left(\mtx{X}_{\ell}\vct{\theta}_\tau-\vct{y}_{\ell}\right).
\end{align*}
The master then aggregates all of these partial gradients and performs the following update
\begin{align}
\label{originaliters}
\vct{\theta}_{\tau+1}=&\mathcal{P}\left(\vct{\theta}_\tau-\mu_\tau\sum_{\ell=1}^L\nabla \mathcal{L}^{(\ell)}(\vct{\theta}_\tau)\right)\nonumber\\
=&\mathcal{P}\left(\vct{\theta}_\tau-\mu_\tau\sum_{\ell=1}^L\mtx{X}^T_{\ell}\left(\mtx{X}_{\ell}\vct{\theta}_\tau-\vct{y}_{\ell}\right)\right)\nonumber\\
=&\mathcal{P}\left(\vct{\theta}_\tau-\tilde{\mu}_\tau\mtx{X}^T(\mtx{X}\vct{\theta}_\tau-\vct{y})\right).
\end{align}
Here, $\mathcal{P}$ denotes the Euclidean projection onto the constraint set $\mathcal{K}=\{\vct{\theta}\in\R^d: \mathcal{R}(\vct{\theta})\le R\}$ and $\tilde{\mu}_\tau$ is the learning rate. 

As mentioned in Section~\ref{sec:intro}, a major performance bottleneck that arises when implementing such distributed PGD updates is that some of the worker nodes may run slow (i.e., stragglers). The presence of such stragglers can significantly slow down the computations as the master has to wait for all the workers to send their partial gradient calculations, so that the overall run time is limited by the slowest worker. For instance, in \cite{anan2010} the effect of slow workers under the title of \textit{outliers} was studied and it was shown that completion  time of jobs can be prolonged by \%34 at median. In this paper, we will focus on the \emph{encoded optimization framework}, which will be described next, to mitigate the effect of stragglers.

\subsection{The Encoded Optimization Framework}\label{EncOpt}

To deal with the effect of stragglers in the iterations \eqref{originaliters}, in this paper we utilize a new approach for straggler mitigation, named \emph{encoded distributed optimization} \cite{Karakus,Karakus-NIPS}
%\MS{Mohammadreza pls put the correct reference here}
, which was originally developed for unconstrained least-squares problems. The main idea behind this approach is to create redundancy in the data by random embedding/encoding. In this section we discuss this computational paradigm tailored to distributed PGD iterates. 

To overcome the computational slowdown caused by stragglers we randomly embed/encode the data by multiplying the feature matrix and the response vector by an encoding matrix $\mtx{A}\in\R^{m\times n}$. We then partition these embedded data $\mtx{A}\mtx{X}/\mtx{A}\vct{y}$ and then distribute them across rows between the $L$ worker nodes. Specifically, we partition the matrix $\mtx{A}$ into $L$ parts across rows $\mtx{A}=\begin{bmatrix}\mtx{A}_1^T & \mtx{A}_2^T & \ldots & \mtx{A}_L^T\end{bmatrix}^T$ with $\mtx{A}_\ell\in\R^{n_\ell\times d}$ and $n=\sum_\ell n_\ell$. Similar to the un-coded case, with this partition worker $\ell$ receives data $\mtx{A}_\ell\mtx{X}/\mtx{A}_\ell\vct{y}$ and carries out updates/computations on this data so that the partial gradient updates are now calculated based on these randomly encoded data. That is,
\begin{align*}
\nabla \mathcal{L}^{(\ell)}(\vct{\theta})=\mtx{X}^T\mtx{A}_\ell^T\mtx{A}_\ell\left(\mtx{X}\vct{\theta}_\tau-\vct{y}\right).
\end{align*}
Due to the effect of stragglers the master may not receive gradient updates from some workers in a timely manner so that computations based on some of the $m$ rows maybe missing. Let us denote the index of the slow workers at iteration $\tau$ by $\mathcal{I}_\tau\subset\{1,2,\ldots,L\}$. Also let $\mathcal{W}_\ell\subset \{1,2,\ldots,n\}$ denote the index of the rows of $\mtx{A}$ sent to worker $\ell$. Now define 

\begin{align*}
\mathcal{S}_\tau=\underset{\ell\in\mathcal{I}_\tau}{\bigcup}\mathcal{W}_\ell,
\end{align*}
which contains the indices of all the rows of $\mtx{A}$ that is not available at the master due to stragglers. We will use $s_\tau=\abs{\mathcal{S}_\tau}$ to denote the total number of these straggler rows.

Based on the gradient updates available to the master it proceeds with the following PGD update
\begin{align}
\label{encodediters}
\vct{\theta}_{\tau+1}=&\mathcal{P}\left(\vct{\theta}_\tau-\mu_\tau\sum_{\ell\in\mathcal{I}_\tau^c}\nabla \mathcal{L}^{(\ell)}(\vct{\theta}_\tau)\right)\nonumber\\
%=&\mathcal{P}\left(\vct{\theta}_\tau-\mu_\tau\mtx{X}^T\left(\sum_{\ell\in\mathcal{I}_\tau^c}\mtx{A}_\ell^T\mtx{A}_\ell\right)(\mtx{X}\vct{\theta}_\tau-\vct{y})\right)\nonumber\\
=&\mathcal{P}\left(\vct{\theta}_\tau-\mu_\tau \mtx{X}^T\mtx{A}_{\mathcal{S}_{\tau}^c}^T\mtx{A}_{\mathcal{S}_{\tau}^c}(\mtx{X}\vct{\theta}_\tau-\vct{y})\right).
\end{align}
Therefore the master effectively runs the encoded iterations \eqref{encodediters} in lieu of the uncoded iterates \eqref{originaliters}. Note that, apriori it is not clear when/why the encoded iterates serve as a good proxy for the uncoded ones. Understanding this relationship is the main focus of this paper. In the next section we discuss the main problems that we study in this paper by formalizing various fundamental trade-offs that arise in the distributed encoded optimization framework.

%In this paper our purpose is to analyze and understand the performance of encoded distributed optimization in a distributed setup, specifically the iterations in \eqref{newiters}, and compare to the cases in which there is no redundancy in data or we replicate the data among the workers.\\

%The question is how does these two iterative updates compare to each other. In this paper we aim to  understand how much closer our iterates become to $\vct{\theta}^*$ after each iteration assuming an adversarial choice of the straggler patterns $\mathcal{S}_\tau$. 

\subsection{Fundamental Resource Trade-offs}\label{trade}

In this paper we wish to understand under what conditions the encoded iterates \eqref{encodediters} are a good proxy for the uncoded iterates \eqref{originaliters}. We aim to answer fundamental questions such as: When will both set of iterates converge to the same fixed point? How does the convergence behavior change due to the presence of the encoding mapping? What are the various trade-offs involved between various resources. To discuss these problems more precisely we start with two simple definitions related to the encoding matrix $\mtx{A}$.

\begin{definition} [Computational Load] We use computational load $m$ to refer to the number of rows of $\mtx{A}$.
\end{definition}

\begin{definition} [Straggler Toleration] \label{stagdef} We use $s$ to denote the maximum number of straggler rows in each iteration ($\abs{\mathcal{S}_\tau}\le s$). We refer to this quantity as straggler toleration.
\end{definition}

With these definitions in place, we now discuss the fundamental trade-offs that we aim to characterize in this paper.

%We wish to understand the converge behavior of the encoded iterates to the optimal solution of \eqref{original}. Stated differently, we wish to understand the convergence behavior to the true parameter ($\vct{\theta}^*$) when there is no noise in the model or an approximate neighborhood (in the presence of noise). Furthermore, i

\begin{itemize}
\item \textbf{Accuracy vs. Computational Time:} In many modern learning applications we must operate on a fixed time budget. Therefore, it is crucial to understand how many passes (or iterations) of the algorithm is required to get to a certain accuracy. We wish to characterize this fundamental trade-off between computational time and accuracy for the encoded distributed optimization framework.  Stated more formally, we are interested in precisely understanding the distance between the encoded iterates and the true parameter ($\twonorm{\vct{\theta}_\tau-\vct{\theta}^*}$) as a function of the number of iterations ($\tau$) and the noise level $\twonorm{\vct{w}}$. 
%how much closer the approximation solution becomes to $\vct{\theta}^{*}$ after $\tau$ iterations using \eqref{newiters}. Specifically, we aim to find the relationship between $\twonorm{\vct{\theta}_\tau-\vct{\theta}^*}$ (distance of approximate solution to original solution) and the noise $\twonorm{\vct{w}}$. %It will be shown that by adjusting the size of matrix $\mtx{A}$, which controls the computation load of iterations, and the number of workers for which the master waits, one can obtain a good approximation of the original solution in a few number of iterations.\\

\item \textbf{Convergence Rate vs. Computational Load:} We are interested in understanding how the computational load $m$ affects the convergence behavior of the encoded iterates. Intuitively, as the computational load increases the encoded iterates provide a better approximation to the un-coded iterates. Therefore, we expect the coded iterates to converge faster as the computational load increases. We wish to precisely characterize the convergence rate as a function of the computational load.

\item \textbf{Convergence Rate vs. Straggler Toleration:} In each encoded iteration, there are some stragglers which are ignored by the master node. We aim to characterize the impact of stragglers on the speed of convergence. By increasing the number of stragglers ($s$), the master node ignores more and more data. Therefore, intuitively we expect that the more stragglers we have, the more iterations are needed for the encoded iterates to converge to a certain accuracy. We wish to characterize the convergence rate as a function of the straggler toleration parameter.%in order that $\vct{\theta}_{\tau} $ closes sufficiently to the original solution $\vct{\theta}^{*}$.

\item \textbf{Computational Load vs. Straggler Toleration:} Intuitively, as we increase the number of stragglers, $s$,  we need more redundancy in our encoded framework. Stated differently, we need to increase the computational load as a function of the number of stragglers. This leads to a fundamental trade-off between computational load and straggler toleration. We aim to characterize the minimum required computational load as a function of the straggler toleration parameter so as to ensure the encoded iterates eventually converge to a good estimate.
\end{itemize}

In the next section we state our main result that leads to a precise characterization of the convergence behavior of the encoded iterates as a function of various parameters, allowing us to precisely characterize the above trade-offs.

%However, this question arises that what is the relationship between computational load, rate of convergence, and the number of stragglers which can be tolerated in each iteration.more precisely, If we denote $m$, the number of rows of the matrix $\mtx{A}$, as the computation load of iterations, we are interested in knowing the fact that for a given matrix $\mtx{A}$ and hence for a given computation load, how many stragglers can be tolerated in each iteration. In addition, For a given  number of stragglers can be tolerated in each iteration, what is the minimum number of computation load ,$m$, to achieve a close approximation of original solution.\\
%Furthermore, we want to characterize the distance of the approximate solution in each iteration of \eqref{newiters}, $\vct{\theta}_{\tau}$, to the original solution , $\vct{\theta}^{*}$, in terms of computation load and number of stragglers can be tolerated in each iteration.\\

%In the following, we characterize the tradeoffs between the rate of convergence of iterations in \eqref{newiters} and the computation load, which is the the number of rows of matrix $\mtx{A}$ as well as the number of stragglers which can be tolerated in each iteration. Additionally, we provide numerical results depicting these tradeoffs clearly.

\section{Main Results}

We wish to characterize the convergence behavior of the encoded iterates \eqref{originaliters} as a function of various problem parameters for the worse possible choice of $s$ straggler rows. More precisely, we are interested in characterizing the relationship between

\begin{align*}
\underset{\mathcal{S}_\tau\subset \{1,2,\ldots,m\},\text{ }\abs{\mathcal{S}_\tau}\le s}{\sup}\twonorm{\vct{\theta}_\tau-\vct{\theta}^*},
\end{align*}
and the error $\twonorm{\vct{w}}$ when running the iterations \eqref{originaliters}. To make these connections precise and quantitative we need a few definitions. 

Naturally our results depend on how well the regularization function $\mathcal{R}$ can capture the properties of the unknown parameter $\vct{\theta}^*$. For example, if we know our unknown parameter is approximately sparse, then using an $\ell_1$ norm for the regularizer is superior to using an $\ell_2$ regularizer. To quantify this capability we first need a couple of standard definitions which we adapt from \cite{Oymak:2015aa, oymak2016fast}.

\begin{definition}[Descent Set and Cone] \label{decsetcone} The \emph{set of descent} of  a function $\mathcal{R}$ at a point $\vct{\theta}^*$ is defined as
\begin{align*}%\text{for some } c\geq 0
{\cal D}_{\mathcal{R}}(\vct{\theta}^*)=\Big\{\vct{h}:\text{ }\mathcal{R}(\vct{\theta}^*+\vct{h})\le \mathcal{R}(\vct{\theta}^*)\Big\}.
\end{align*}
The \emph{cone of descent} is defined as a closed cone $\mathcal{C}_{\mathcal{R}}(\vct{\theta}^*)$ that contains the descent set, i.e.~$\mathcal{D}_{\mathcal{R}}(\vct{\theta}^*)\subset\mathcal{C}_{\mathcal{R}}(\vct{\theta}^*)$. The \emph{tangent cone} is the conic hull of the descent set. That is, the smallest closed cone $\mathcal{C}_{\mathcal{R}}(\vct{\theta}^*)$ obeying $\mathcal{D}_{\mathcal{R}}(\vct{\theta}^*)\subset\mathcal{C}_{\mathcal{R}}(\vct{\theta}^*)$.
\end{definition}

We note that the capability of the regularizer $\mathcal{R}$ in capturing the properties of the parameter vector $\vct{\theta}^*$ depends on the size of the descent cone $\mathcal{C}_{\mathcal{R}}(\vct{\theta}^*)$. The smaller this cone is the more suited the function $\mathcal{R}$ is at capturing the properties of $\vct{\theta}^*$. To quantify the size of various cones we shall use the notion of mean width.

\begin{definition}[Gaussian Width]\label{Gausswidth} The Gaussian width of a set $\mathcal{C}\in\R^n$ is defined as $\omega(\mathcal{C}):=\mathbb{E}_{\vct{g}}[\underset{\vct{z}\in\mathcal{C}}{\sup}~\langle \vct{g},\vct{z}\rangle]$, where the expectation is taken over $\vct{g}\sim\mathcal{N}(\vct{0},\mtx{I}_n)$. 
\end{definition}

We now have all the definitions in place to quantify the capability of the function $\mathcal{R}$ in capturing the properties of the unknown parameter $\vct{\theta}^*$ when using an encoding matrix $\mtx{A}$. This naturally leads us to the definition of the minimum required computational load.

\begin{definition}[Minimal Computational Load]\label{PTcurve}
Let $\mathcal{C}_{\mathcal{R}}(\vct{\theta}^*)$ be a cone of descent of $\mathcal{R}$ at $\vct{\theta}^*$. We define the minimal computational load function as
\begin{align*}
\mathcal{M}(\mathcal{R},\mtx{X},\vct{\theta}^*,\eta)=\left(\omega\left(\mtx{X}\mathcal{C}_{\mathcal{R}}(\vct{\theta}^*)\cap\mathbb{S}^{n-1}\right)+\eta\right)^2.
\end{align*}
where $\vct{\theta}^{*}\in\R^{d}$ and $\mtx{X}\in\R^{n\times d}$. We shall often use the short hand $m_0=\mathcal{M}(\mathcal{R},\mtx{X},\vct{\theta}^*,\eta)$ with the dependence on $\mathcal{R}, \mtx{X}, \vct{\theta}^*, \eta$ implied.  
\end{definition}

The definition above characterizes the minimum computational load required for the encoded iterations to converge to the true parameter in the absence of noise or stragglers.

The convergence rate of the encoded iterates also naturally depends on various characteristics of the feature matrix $\mtx{X}$. We quantify a few of these characteristics below.

\begin{definition}[Cone-Restricted Spectral Norm]\label{minconeig} Let $\mathcal{C}_{\mathcal{R}}(\vct{\theta}^*)$ be the cone of descent of the regularization function $\mathcal{R}$ at a point $\vct{\theta}^*$ per Definition \ref{decsetcone}. the cone-restricted spectral norm of a matrix $\mtx{X}\in\R^{n\times d}$ with respect to $\mathcal{R}$ at a point $\vct{\theta}^*$ is defined as $\sigma_{\mathcal{R}}(\mtx{X})=\underset{\vct{u}\in\mathcal{C}_{\mathcal{R}}(\vct{\theta}^*)\cap\mathbb{S}^{d-1}}{\sup}\twonorm{\mtx{X}\vct{u}}$.
\end{definition}

We note that the above definition is a natural extension of the spectral norm of a matrix. It is well known that the spectral norm of the feature matrix plays a crucial role in the convergence behavior of least-square problems. The cone-restricted spectral norm defined above plays a similar role in the convergence of constrained least-squares problems.

Furthermore, the convergence behavior of the encoded iterates is also related to that of the uncoded iterates. The following two definitions, adapted from \cite{Oymak:2015aa}, concern the convergence of the uncoded iterates.

\begin{definition}[Convergence Rate]\label{convrate}
Consider the iterations \ref{originaliters}. Let $\vct{\theta}^{*}\in\R^{d}$ and $\mtx{X}\in\R^{n\times d}$ and $\mathcal{R}$ be the regularizer function as well as $\mu$ be the learning rate. We define 
{\small
\begin{align*}
\rho(\mu):=\rho(\mu,\mtx{X},\mathcal{R},\vct{\theta}^*)=\underset{\vct{u},\vct{v}\in\mathcal{C}_{\mathcal{R}}(\vct{\theta}^*)\cap \mathbb{S}^{d-1}}{\sup}\vct{u}^T\left(\mtx{I}-\mu\mtx{X}^T\mtx{X}\right)\vct{v},
\end{align*}
}
\end{definition}

It is known that $\rho(\mu)$ characterizes the convergence rate of the uncoded iterations \eqref{originaliters} \cite{Oymak:2015aa}.

\begin{definition}[Noise Amplification]\label{noisamp}
Consider the iterations \ref{originaliters}. Let $\vct{\theta}^{*}\in\R^{d}$ and $\mtx{X}\in\R^{n\times d}$ and $\mathcal{R}$ be the regularizer function with $\vct{w}$ denoting the noise.
We define
{\small
\begin{align*}
\xi(\mtx{X}):=\xi(\mtx{X},\mathcal{R},\vct{\theta}^*,\vct{w})=\underset{\vct{v}\in-\mathcal{C}_{\mathcal{R}}(\vct{\theta}^*)\cap\mathbb{S}^{d-1}}{\sup}\vct{v}^T\mtx{X}^T\frac{\vct{w}}{\twonorm{\vct{w}}}.
\end{align*}
}
\end{definition}

It is known that the uncoded iterations eventually converge to a neighborhood of the unknown parameter $\vct{\theta}^{*}$ \cite{Oymak:2015aa}. The noise amplification factor defined above plays a crucial role in characterizing the size of this neighborhood. In particular, \cite{Oymak:2015aa} shows that the diameter of this neighborhood is proportional to $\xi(\mtx{X})\twonorm{\vct{w}}$.

With these definitions in place we are now ready to state our main theorem regarding the convergence of the encoded iterates \eqref{encodediters}.
\begin{theorem}\label{mainTheom} Let $\mtx{A}\in\R^{m\times n}$ be a matrix with i.i.d.~$\mathcal{N}(0,1)$ entries. Also assume the number of straggler rows obeys $s_\tau=\abs{\mathcal{S}_\tau}\le s$ with $s$ the straggler toleration parameter per Definition \ref{stagdef} obeying $s\le m$. Furthermore, let $m_0=\mathcal{M}(\mathcal{R},\mtx{X},\vct{\theta}^*,\eta)$ denote the minimal computational load per Definition \ref{PTcurve}. 
%assume 
%\begin{align}
%\label{samp}
%(m-s_\tau)\ge c m_0,
%\end{align}
%\begin{align}
%\label{samp}
%4(m-s_\tau)\log\left(\frac{em}{m-s_\tau}\right)\ge cm_0,
%\end{align}
%with
Then the encoded iterative updates \eqref{encodediters} obey 
\begin{align}
\label{ineqbnd}
\underset{\mathcal{S}_{\tau}\subset \{1,2,\ldots,m\}}{\sup}\twonorm{\vct{\theta}_{\tau+1}-\vct{\theta}^*}\nonumber
\le& \kappa_{\mathcal{R}}\rho(\tilde{\mu}_\tau)\twonorm{\vct{\theta}_{\tau}-\vct{\theta}^*}\nonumber\\
&+\tilde{\mu}_\tau\cdot\kappa_{\mathcal{R}}\cdot\sigma_{\mathcal{R}}^{2}(\mtx{X})\cdot\nonumber
\left(\frac{2+9s_\tau\log(em/s_\tau)}{m}+4\sqrt{\frac{m_0}{m-s_\tau}}\right)\twonorm{\vct{\theta}_{\tau}-\vct{\theta}^*}\nonumber\\
&+\kappa_{\mathcal{R}}\bigg(\tilde{\mu}_{\tau}\cdot\xi(\mtx{X})+\frac{\tilde{\mu}_\tau}{\sqrt{2}}\cdot\sigma_{\mathcal{R}}(\mtx{X})\sqrt{\frac{m_0}{m-s_\tau}}\bigg) \twonorm{\vct{w}},
\end{align}
for all $\tau$ with probability at least $1-6e^{-\frac{\eta^2}{8}}-e^{-\frac{\eta^2}{2}}-e^{-\frac{m}{2}}$. Here, 
$\rho$ is the convergence rate per Definition \ref{convrate}, $\xi$ is the noise amplifications per Definition \ref{noisamp}, $\sigma_{\mathcal{R}}(\mtx{X})$ is the cone-restricted spectral norm of $\mtx{X}$ per Definition \ref{minconeig}.  Furthermore, the tunning parameter is set to $R=\mathcal{R}(\vct{\theta}^*)$ and the learning rate is equal to $\mu_\tau=\frac{\tilde{\mu}_\tau}{\beta_{s_\tau,m}^2}$ with $\beta_{s,m}=\min(\sqrt{3(m-s)\log\left(\frac{em}{m-s}\right)},\sqrt{m})$ and $\tilde{\mu}_\tau$ is the learning rate in the uncoded iterations \eqref{originaliters}. Finally, $\kappa_{\mathcal{R}}=1$ for convex $\mathcal{R}$ and $\kappa_{\mathcal{R}}=2$ for nonconvex $\mathcal{R}$.
%Throughout we use $\mathcal{S}^{d-1}$ to denote the unit sphere of $\R^d$.
%Also, define
%\begin{align*}
%\rho(\mu):&=\rho(\mu,\mtx{X},\mathcal{R},\vct{\theta}^*)\\&=\underset{\vct{u},\vct{v}\in\mathcal{C}_{\mathcal{R}}(\vct{\theta}^*)\cap \mathbb{S}^{d-1}}{\sup}\vct{u}^T\left(\mtx{I}-\mu\mtx{X}^T\mtx{X}\right)\vct{v},
%\end{align*}
%and
%\begin{align*}
%\xi(\mtx{X}):&=\xi(\mtx{X},\mathcal{R},\vct{\theta}^*,\vct{w})\\&=\underset{\vct{v}\in-\mathcal{C}_{\mathcal{R}}(\vct{\theta}^*)\cap\mathbb{S}^{d-1}}{\sup}\vct{v}^T\mtx{X}^T\frac{\vct{w}}{\twonorm{\vct{w}}}.
%\end{align*}

%$\rho(\tilde{\mu}_\tau)$ and $\xi(\mtx{X})$ be the convergence rate and noise amplification factor per Theorem \ref{deterministic}. 
%Then the iterative updates \eqref{newiters} obey 
%\begin{align*}
%&\underset{\mathcal{S}_{\tau}\subset \{1,2,\ldots,m\}}{\sup}\twonorm{\vct{\theta}_{\tau+1}-\vct{\theta}^*}\\&\le\kappa_{\mathcal{R}}\left(\rho(\tilde{\mu}_\tau)+\frac{3}{\sqrt{c}}\cdot\tilde{\mu}_\tau\cdot\sigma_{\mathcal{R}}^{2}(\mtx{X})\right)\twonorm{\vct{\theta}_{\tau}-\vct{\theta}^*}\\
%&+\kappa_{\mathcal{R}}\left(\frac{1}{2\log\left(\frac{m}{m-s_\tau}\right)}\tilde{\mu}_{\tau}\cdot\xi(\mtx{X})+\tilde{\mu}_\tau\sqrt{\frac{m_0}{16(m-s_\tau)\log^2\left(\frac{m}{m-s_\tau}\right)}}\right)\twonorm{\vct{w}},
%\end{align*}
%for all $\tau$ with probability at least $1-6e^{-\frac{\eta^2}{8}}-e^{-\frac{\eta^2}{2}}-e^{-\frac{m}{2}}$.
%Throughout we use $\mathcal{S}^{d-1}$ to denote the unit sphere of $\R^d$.
\end{theorem}

\begin{remarks}
Theorem \ref{mainTheom} essentially connects the convergence behavior of the encoded iterations to that of the uncoded iterations. Consider the limit $m\rightarrow \infty$ and note that for a Gaussian matrix $\mtx{A}_{\mathcal{S}_\tau^c}$, $\mtx{A}_{\mathcal{S}_\tau^c}^T\mtx{A}_{\mathcal{S}_\tau^c}\rightarrow\mtx{I}$ and thus the encoded iterations reduce to the uncoded iterations. In this case the convergence bound provided by Theorem \ref{mainTheom} reduces to
{\small
\begin{align}
\label{uncodedg}
\twonorm{\vct{\theta}_{\tau+1}-\vct{\theta}^*}\le \kappa_{\mathcal{R}}\rho(\tilde{\mu}_\tau)&\twonorm{\vct{\theta}_\tau-\vct{\theta}^*}+\frac{\kappa_{\mathcal{R}}}{\sqrt{2}}\tilde{\mu}_\tau\cdot \xi(\mtx{X}))\twonorm{\vct{w}}.
\end{align}
}
The first term gives the convergence rate to the true parameter. The second term characterizes the size of the neighborhood (of the true parameter) to which the iterates converge, demonstrating that the iterates eventually approximate the true parameter up to a term that is proportional to the Euclidean norm of the noise. The bound \eqref{uncodedg} was proven recently in \cite{Oymak:2015aa}%\MS{Mohammadreza pls provide the right citation}.
. Our result generalizes this result to the encoded case while recovering the special uncoded case in the limit $m\rightarrow \infty$.  
\end{remarks}
\vspace{-0.4cm}
\begin{remarks}
Theorem \ref{mainTheom} characterizes the minimal computational load for convergence of \eqref{encodediters} in the presence of stragglers. Comparing \eqref{ineqbnd} with \eqref{uncodedg} we see that as long as the computational load is sufficiently large the effect of coding is only a slight increase in the convergence rate and the size of approximation neighborhood. For instance, to ensure that in the encoded case the convergence rate only increases by $\epsilon$ the computational load must obey
\begin{align*}
m-s\ge 260\frac{m_0+s}{\epsilon^2}\ge 64 \frac{m_0+s}{\epsilon^2}\kappa_{\mathcal{R}}^2 (\tilde{\mu}_\tau\cdot \sigma_{\mathcal{R}}^2(\mtx{X}))^2.
\end{align*}
In the last inequality we used $\kappa_{\mathcal{R}}\le 2$ and the fact that $\tilde{\mu}_\tau$ typically scales with $1/\sigma_{\mathcal{R}}(\mtx{X})^2$ (as step size typically scales with the inverse of the smoothness parameter). Thus as long as the computational load exceeds the sum of the number of stragglers and the minimal computational load by a constant factor, i.e.~
\begin{align}
\label{samp}
    m \ge c(m_0+s)
    \end{align}
holds for some numerical constant $c$ depending only on $\epsilon$, then the increase in the convergence rate is small. Similarly, the increase in the size of the approximation neighborhood remains small as long as \eqref{samp} holds.

\end{remarks}

%\begin{remarks}
%Comparing the result of Theorem \ref{mainTheom} to that of Theorem 1.2 in \cite{Oymak:2015aa}, in which the iterations \eqref{originaliters} is used, we can see the price of ignoring stragglers for computing the gradients in each iteration in the extra term added which decreases the rate of convergence up to the factor   $\frac{3}{\sqrt{c}}\cdot\tilde{\mu}_\tau\cdot\sigma_{\mathcal{R}}^{2}(\mtx{X})$. Moreover, the iterations in \eqref{originaliters} and \eqref{newiters} guarantee that approximation solution lies in the neighborhood of the original solution such that the radius of the neighborhood is proportional to the magnitude of noise. In the Theorem \ref{mainTheom}, the radius is larger than that in Theorem 1.2 in \cite{Oymak:2015aa} which shows the effect of stragglers.
%\end{remarks}
%Now we intend to investigate the resource tradeoffs explained in section \ref{trade}.
\begin{remarks}
We now briefly discuss how our results compare with related work. Theorem \ref{mainTheom} demonstrates that the encoded iterates converge at a linear rate whilst dealing with arbitrary and possibly nonconvex constraints. \cite{Karakus-NIPS} also demonstrates a linear convergence, albeit in terms of the optimal value. However, \cite{Karakus-NIPS} only focuses on the special case where there are no constraints on the optimization variables. Furthermore, \cite{Karakus-NIPS} requires a computational load that is larger than the sum of the number of stragglers and the total number of data points i.e.~$m\ge 2(n+s)$. In comparison, our results require a near minimal number of samples that is commensurate to the sum of the straggler toleration and the amount of prior knowledge ($m\ge c(m_0+s)$). This allows for a much smaller computation load that can even be significantly smaller than the number of data points i.e.~$m<<n$. Finally, we would like to mention related work in \cite{Pilanci} where the authors focus on sketching of constrained convex programs. This paper focuses on the properties of the optimal solution to problems of the form \eqref{original} without any stragglers. In comparison, we focus on analyzing the convergence behavior of iterative algorithms when stragglers are present.
%It is also worthwhile mentioning that in \cite{Pilanci} the problem of data reduction for constrained convex programs is studied. However, we give a convergence guarantee for constrained convex programs with low computational load in the presence of stragglers.
%In comparison, the results Furthermore, Theorem \ref{mainTheom} guarantees that as long as \eqref{samp} is satisfied it achieves convergence. This means that with a small size of computational load which can even be so much less than the number of data points the approximate solution converges to the original one, while in \cite{Karakus-NIPS} the redundancy factor has to be almost $2$ resulting in too much computational load. 
\end{remarks}

\begin{remarks}[Convergence Rate vs. Computational Load]
Theorem \ref{mainTheom} characterizes the effect of the computational load on the convergence rate. In particular, this theorem shows that the increase in the convergence rate is proportional to $1/\sqrt{m}$. Therefore, as the computational load increases the convergence rate decreases. Thus, a larger computational load ensures a faster convergence of the encoded iterates.
\end{remarks}

\begin{remarks}[Convergence Rate vs. Straggler Toleration]
Theorem \ref{mainTheom} also characterizes the effect of
stragglers on the rate of convergence. This result demonstrates a rate proportional to $1/\sqrt{m-s}$ so that as the number of stragglers increase, the convergence rate decreases leading to a slower convergence of the encoded iterates. 
\end{remarks}

%%Increasing $$, by increasing $s_\tau$, the constant $c$ in \eqref{samp} would be decreased which implies increasing in the rate of convergence as well.
\begin{remarks}[Computational Load vs. Straggler Toleration]
We also note that Theorem \ref{mainTheom} indirectly characterizes a trade-off between the computational load and the straggler toleration of the encoded iterations through \eqref{samp}. Indeed, \eqref{samp} demonstrates that for a fixed convergence rate the computational load must scale linearly with the number of stragglers.
\end{remarks}

\section{Numerical Results}

In this section, we corroborate the resource trade-offs characterized in Theorem \ref{mainTheom} via experiments on synthetic data. We generate the true parameter $\vct{\theta}^*\in\R^{d}$ with $d=4000$ and sparsity level $k=20$ where the support is chosen at random and the values on support are distributed i.i.d $N(0,1)$. Moreover, we generate the data matrix $\mtx{X}\in\R^{n\times d}$ i.i.d. $\sim N(0,1)$ with $n=3000$ and set the output vector via $\vct{y}=\mtx{X}\vct{\theta}^*$. In our simulations we vary the computational load $m$ and the straggler toleration $s$ and then plot the various trade-offs. We use two different encoding matrices: a random Gaussian matrix and a random Discrete Cosine Transform (DCT) matrix. In the Gaussian case the entries of the matrix are generated i.i.d.~$\sim N(0,1)$. The random DCT matrix is generated according to $\mtx{A}=\mtx{H}\mtx{D}$ where $\mtx{H}\in\R^{m\times n}$ is obtained by selecting $m$ rows of an $n\times n$ DCT matrix at random and $\mtx{D}\in\R^{n\times n}$ is a diagonal matrix with i.i.d.~$\pm 1$ entries on the diagonal. In our simulations in each iteration we assume a different set of straggler rows chosen i.i.d.~at random from the $m$ rows. To reconstruct $\vct{\theta}^*$, we run encoded Projected Gradient Descent (PGD) iterations \eqref{encodediters} for solving \eqref{original} with learning rates $\mu_\tau=\frac{1}{5m}$ and $\mu_\tau=1/3$ for the Gaussian and randomized DCT encoding matrices, respectively. We use $\mathcal{R}(\vct{\theta})=\onenorm{\vct{\theta}}$ with tuning parameter $R=||\vct{\theta}^*||_1$. We run the encoded PGD iterates for $500$ iterations and record the relative error $\twonorm{\vct{\theta}_\tau-\vct{\theta}^*}/ {\twonorm{\vct{\theta}^*}}$.
%and record the empirical probability of success. We take average over 50 trials and a trial is success if the relative error $\twonorm{\vct{\theta}_\tau-\vct{\theta}^*}/ {\twonorm{\vct{\theta}^*}}$ after $500$ iterations is less than $10^{-3}$, otherwise it would be a failure.
%\begin{figure}[htbp]
 % \centering
  %\includegraphics[width=0.44\textwidth]{plotm.eps}
  %\caption{Plotting the relative error $\frac{\twonorm{\vct{\theta}_\tau-\vct{\theta}^*}}{\twonorm{\vct{\theta}^*}}$ verses the number of iterations using a semi-log graph when fixing $s=100$, where $s$ is straggler toleration, and $m=200,400,600,800$, where $m$ is computational load.}
  %\label{fig:fix_s}
  %\vspace{-6mm}
%\end{figure}

\begin{itemize}
\item\textbf{Convergence Rate vs. Computational Load:} In this simulation we fix the straggler toleration at $s=100$ and vary the computation load $m$. We depict the relative error as a function of iterations in Figure \textcolor{darkred}{1a}. This figure confirms that the convergence is indeed linear and increasing $m$ leads to a faster convergence as predicted by Theorem \ref{mainTheom}.

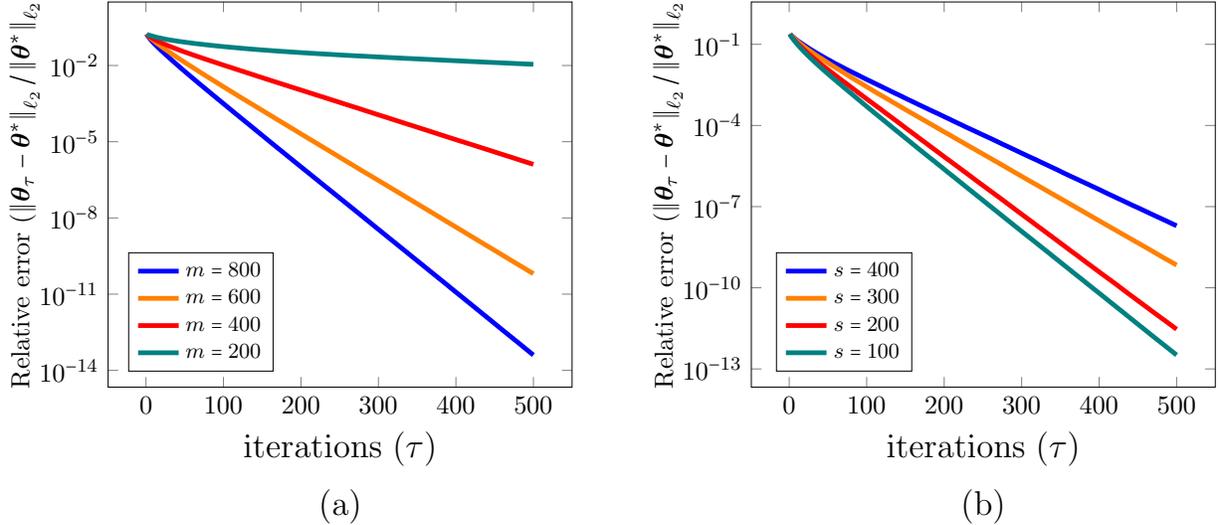
\begin{figure}
\centering
\hspace{-4cm}
    \begin{subfigure}{0.23\textwidth}
    \centering
    \begin{tikzpicture}[scale=.9]
    \begin{semilogyaxis}[xlabel=\huge$\substack{\text{iterations ($\tau$)}\\ \\ \text{(a)}}$\normalsize,
        ylabel=Relative error ($\twonorm{\vct{\theta}_\tau-\vct{\theta}^*}/\twonorm{\vct{\theta}^*}$, legend style={font=\footnotesize,at={(0.20,0.35)},anchor=north,legend cell align=left}]

      \addplot [blue,line width=2pt] table[x index=0,y index=1]{./M1};\addlegendentry{$m=800$}
      
        \addplot [orange,line width=2pt] table[x index=0,y index=1]{./M2};\addlegendentry{$m=600$}
        
       \addplot [red,line width=2pt] table[x index=0,y index=1]{./M3};\addlegendentry{$m=400$}
       
      \addplot [teal,line width=2pt] table[x index=0,y index=1]{./M4};\addlegendentry{$m=200$}    
\end{semilogyaxis}
    
    \end{tikzpicture} 
    \centering
           %\caption{}
            \label{fig1a}
        \end{subfigure} 
        \hspace{4.5cm}
        %\hfill
    \begin{subfigure}{0.23\textwidth}
    \centering
    \begin{tikzpicture}[scale=.9]
    
     \begin{semilogyaxis}[xlabel=\huge$\substack{\text{iterations ($\tau$)}\\ \\ \text{(b)}}$\normalsize,
        ylabel=Relative error ($\twonorm{\vct{\theta}_\tau-\vct{\theta}^*}/\twonorm{\vct{\theta}^*}$, legend style={font=\footnotesize,at={(0.20,0.35)},anchor=north,legend cell align=left}]

      \addplot [blue,line width=2pt] table[x index=0,y index=1]{./S1};\addlegendentry{$s=400$}
      
        \addplot [orange,line width=2pt] table[x index=0,y index=1]{./S2};\addlegendentry{$s=300$}
        
       \addplot [red,line width=2pt] table[x index=0,y index=1]{./S3};\addlegendentry{$s=200$}
       
      \addplot [teal,line width=2pt] table[x index=0,y index=1]{./S4};\addlegendentry{$s=100$}      
\end{semilogyaxis}
    
    \end{tikzpicture}   
            %\caption{}
            \label{fig1b}
        \end{subfigure}
    \vspace{-8mm}    
    \caption{These two diagram show the empirical rates of convergence for Gaussian encoded projected gradient descent in two different scenarios. (a) depicts the converge rate as a function of the computational load $m$ when the straggler toleration is fixed at $s=100$. (b) shows the convergence rate as a function of the straggler toleration with a fixed computational load at $m=800$.}

    \label{fig1}
    \end{figure}

\item\textbf{Convergence Rate vs. Straggler Toleration:}
In this simulation we fix the computational load at $m=800$ and vary the straggler toleration $s$ and depict the relative errors as a function of the iterations in Figure \textcolor{darkred}{1b}. This figure confirms that the iterates converge at a linear rate and increasing $s$ leads to a slower convergence as predicted by Theorem \ref{mainTheom}.

    \begin{figure}[t]
        \centering
        \begin{subfigure}[!t]{0.43\textwidth}
                \centering
                \includegraphics[width=\textwidth]{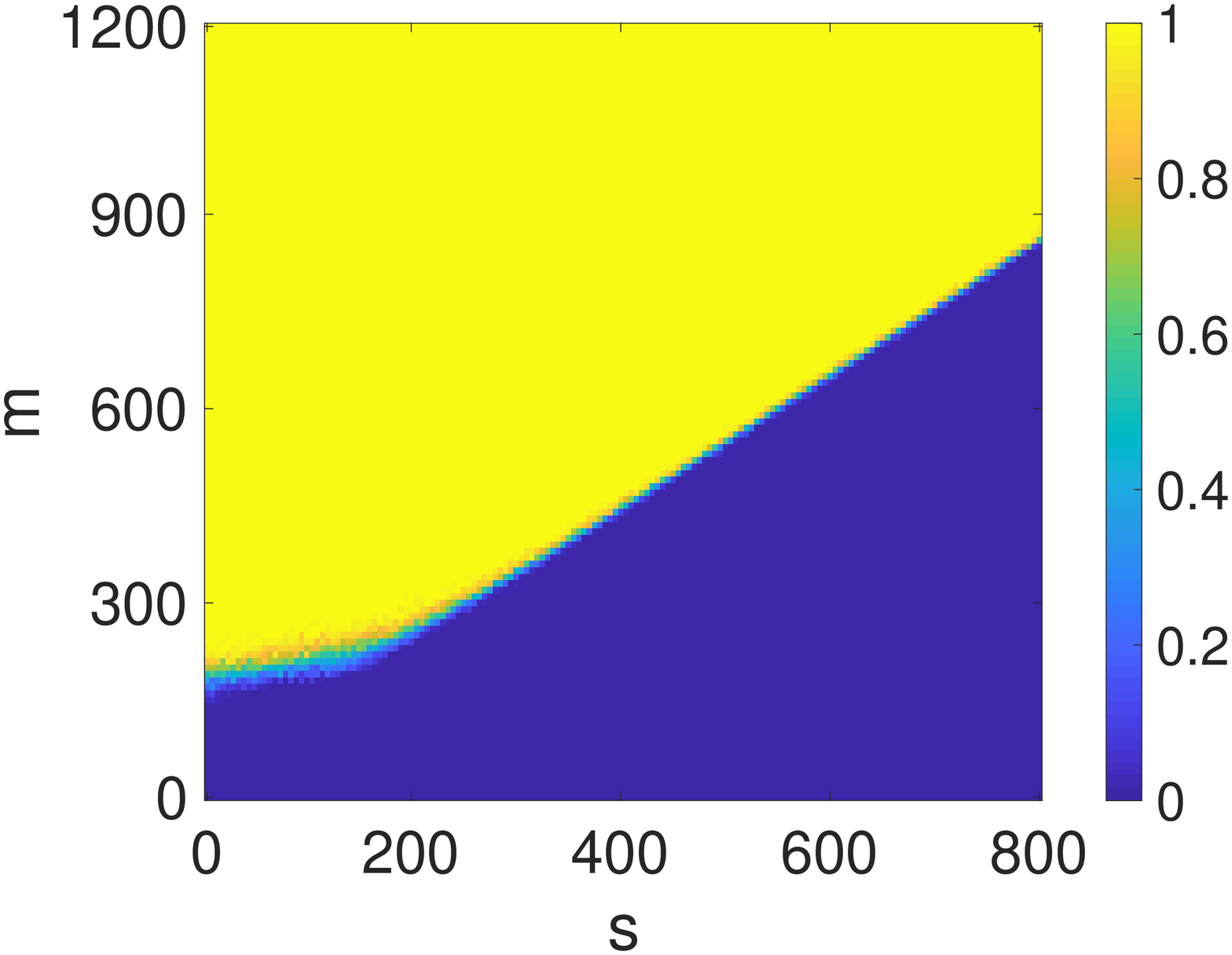}
                \caption{}
                \label{6T_cell}
        \end{subfigure}
        %%%%%%%%%%%%%%%%%%%%%%%%%%
        \begin{subfigure}[!t]{0.47\textwidth}
                \centering
                \includegraphics[width=\textwidth]{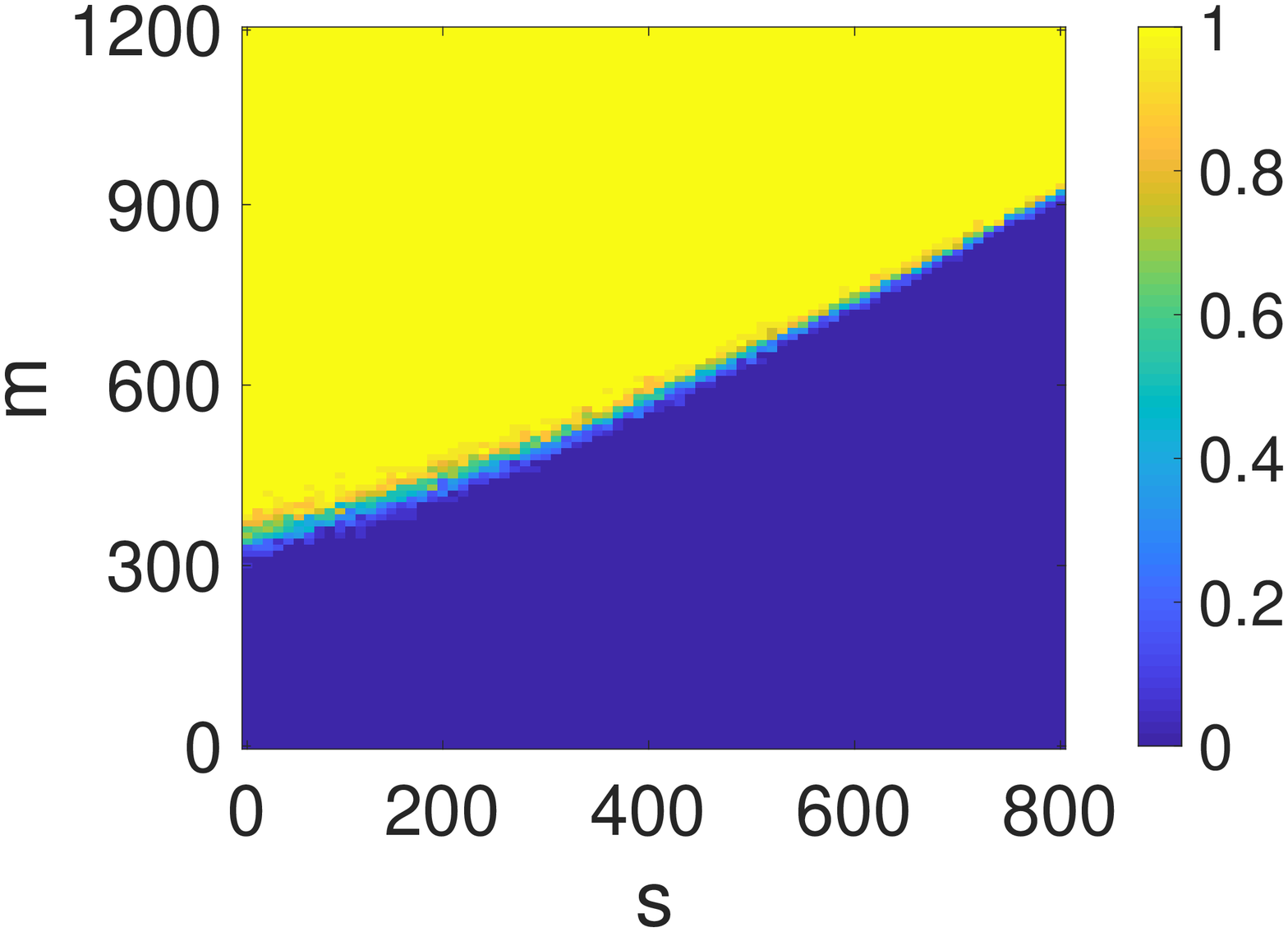}
                \caption{}
                \label{6T_Layout}
        \end{subfigure}
        %%%%%%%%%%%%%%%%%%%%%%%%%%
        
        %%%%%%%%%%%%%%%%%%%%%%%%%%
        \caption{These two diagrams depict the empirical probability that encoded projected gradient descent successfully reaches the global optimum of the uncoded optimization problem for (a) Gaussian and (b) randomized DCT encoding matrices. The colormap tapers between yellow and blue where yellow represents certain success, while blue represents certain failure.}
        \label{Splitter}
\end{figure}

\item\textbf{Computational Load vs. Straggler Toleration:}
In this simulation we vary the computational load $m$, and straggler toleration $s$ and for each case run the encoded PGD iterations. We stop after $500$ iterations and record the empirical probability of success. The empirical probability of success is an average over $50$ trials, where in each instance, we generate new random parameter vectors, data and encoding matrices. We declare a trial successful if the relative error of the reconstruction $\twonorm{\vct{\theta}-\vct{\theta}^*}/\twonorm{\vct{\theta}^*}$ falls below $10^{-3}$.

Figure \textcolor{darkred}{2a} depicts the empirical success probabilities via a color map for different straggler tolerations $s$ and computational loads $m$. Yellow represents certain success, while blue represents certain failure. In the experiments of this figure the encoding matrix is Gaussian. This figure clearly shows that there is a phase transition curve for the computational load as a function of the straggler toleration. On one side
of this curve encoded PGD updates is successful with high probability on the other side it fails with high
probability. Figure \textcolor{darkred}{2a} also shows that the computational load scales linearly in terms of the straggler toleration parameter confirming the relationship \eqref{samp} predicted by Theorem \ref{mainTheom}. Figure \textcolor{darkred}{2b} depicts the results for randomized DCT matrices. Encoding with such matrices is very efficient requiring only a DCT transform. Perhaps unexpectedly, randomized DCT matrices exhibit very similar behavior to the Gaussian matrix demonstrating that such matrices can act as computational friendly surrogates for encoding purposes. Proving Theorem \ref{mainTheom} extends to randomized DCT matrices is an interesting direction for future research.

\end{itemize}

%\section{Acknowledgements}

%This research is in part supported by Northrop Grumman Cybersecurity Research Consortium. A.~S.~is supported by Defense Advanced Research Projects Agency (DARPA) under Contract No. HR001117C0053. The views, opinions, and/or findings expressed are those of the author(s) and should not be interpreted as representing the official views or policies of the Department of Defense or the U.S. Government. A.~S.~is also in part supported by ONR award N000141612189, NSF Grants CCF-1703575 and NeTS-1419632. M.~S.~is supported by the Air Force Office of Scientific Research Young Investigator Program (AFOSR-YIP) under award number FA9550-18-1-0078 and a Google faculty research award. 

\section{Proofs}
\label{proof}

In this section we prove Theorem \ref{mainTheom}. Define the error vector $\vct{h}_\tau:=\vct{\theta}_\tau-\vct{\theta}^*$ and the cones $\widetilde{\mathcal{C}}:=\mathcal{C}_{\mathcal{R}}(\vct{\theta}^*)\in\R^d$ and $\mathcal{C}:=\mtx{X}\widetilde{\mathcal{C}}\in\R^n$. Utilizing \cite{Oymak:2015aa}[Theorem 1.2] we have 
\begin{align}
\label{mainconv}
\twonorm{\vct{h}_{\tau+1}}\le&\kappa_{\mathcal{R}}\bigg(\underset{\tilde{\vct{u}},\tilde{\vct{v}}\in\widetilde{\mathcal{C}}\cap\mathbb{S}^{d-1}}{\sup}\text{ }\tilde{\vct{u}}^T(\mtx{I}-\mu_\tau\mtx{X}^T\mtx{A}_{\mathcal{S}_{\tau}^c}^T\mtx{A}_{\mathcal{S}_{\tau}^c}\mtx{X})\tilde{\vct{v}}\bigg)\twonorm{\vct{h}_\tau}+\kappa_{\mathcal{R}}\cdot\mu_\tau\cdot \underset{\tilde{\vct{u}}\in-\widetilde{\mathcal{C}}\cap\mathbb{S}^{d-1}}{\sup}\text{ }\tilde{\vct{u}}^T\mtx{X}^T\mtx{A}_{\mathcal{S}_{\tau}^c}^T\mtx{A}_{\mathcal{S}_{\tau}^c}\vct{w}.
\end{align}

We now proceed by simplifying each of these two terms. To simplify the first term, define $\vct{u}:=\frac{\mtx{X}\tilde{\vct{u}}}{\twonorm{\mtx{X}\tilde{\vct{u}}}}\in\mathbb{S}^{n-1}$ and $\vct{v}:=\frac{\mtx{X}\tilde{\vct{v}}}{\twonorm{\mtx{X}\tilde{\vct{u}}}}\in\mathbb{S}^{n-1}$ and note that
\begin{align*}
\tilde{\vct{u}}^T(&\mtx{I}-\mu_\tau\mtx{X}^T\mtx{A}_{\mathcal{S}_{\tau}^c}^T\mtx{A}_{\mathcal{S}_{\tau}^c}\mtx{X})\tilde{\vct{v}}
%=&\tilde{\vct{u}}^T\left(\mtx{I}-\frac{\tilde{\mu}_\tau}{\beta_{s_\tau,m}^2}\mtx{X}^T\mtx{A}_{\mathcal{S}_{\tau}^c}^T\mtx{A}_{\mathcal{S}_{\tau}^c}\mtx{X}\right)\tilde{\vct{v}}\\
%=&\tilde{\vct{u}}^T\left(\mtx{I}-\tilde{\mu}_\tau\mtx{X}^T\mtx{X}\right)\tilde{\vct{v}}\\&+\tilde{\mu}_\tau\tilde{\vct{u}}^T\mtx{X}^T\left(\mtx{I}-\frac{1}{\beta_{s_\tau,m}^2}\mtx{A}_{\mathcal{S}_{\tau}^c}^T\mtx{A}_{\mathcal{S}_{\tau}^c}\right)\mtx{X}\tilde{\vct{v}}\\
=\tilde{\vct{u}}^T\left(\mtx{I}-\tilde{\mu}_\tau\mtx{X}^T\mtx{X}\right)\tilde{\vct{v}}+\tilde{\mu}_\tau\cdot\twonorm{\mtx{X}\tilde{\vct{u}}}\cdot\twonorm{\mtx{X}\tilde{\vct{v}}}\vct{u}^T\left(\mtx{I}-\frac{1}{\beta_{s_\tau,m}^2}\mtx{A}_{\mathcal{S}_{\tau}^c}^T\mtx{A}_{\mathcal{S}_{\tau}^c}\right)\vct{v}.
\end{align*}
Now we can use the fact that supremum of sum is less than sum of suprema. Thus,

\begin{align}
\label{temppf1}
\underset{\tilde{\vct{u}},\tilde{\vct{v}}\in\widetilde{\mathcal{C}}\cap\mathbb{S}^{d-1}}{\sup}\text{ }\tilde{\vct{u}}^T(\mtx{I}&-\mu_\tau\mtx{X}^T\mtx{A}_{\mathcal{S}_{\tau}^c}^T\mtx{A}_{\mathcal{S}_{\tau}^c}\mtx{X})\tilde{\vct{v}}\nonumber\\
\le&\underset{\tilde{\vct{u}},\tilde{\vct{v}}\in\widetilde{\mathcal{C}}\cap\mathbb{S}^{d-1}}{\sup}\text{ }\tilde{\vct{u}}^T\left(\mtx{I}-\tilde{\mu}_\tau\mtx{X}^T\mtx{X}\right)\tilde{\vct{v}}\nonumber\\
&+\tilde{\mu}_\tau\Bigg(\underset{\tilde{\vct{u}}\in\widetilde{\mathcal{C}}\cap\mathbb{S}^{d-1}}{\sup}\text{ }\twonorm{\mtx{X}\tilde{\vct{u}}}\cdot\underset{\tilde{\vct{v}}\in\widetilde{\mathcal{C}}\cap\mathbb{S}^{d-1}}{\sup}\text{ }\twonorm{\mtx{X}\tilde{\vct{v}}}\nonumber
\cdot\underset{\vct{u},\vct{v}\in\mathcal{C}\cap\mathbb{S}^{n-1}}{\sup}\text{ }\vct{u}^T\left(\mtx{I}-\frac{1}{\beta_{s_\tau,m}^2}\mtx{A}_{\mathcal{S}_{\tau}^c}^T\mtx{A}_{\mathcal{S}_{\tau}^c}\right)\vct{v}\Bigg)\nonumber\\
=&\rho(\tilde{\mu}_\tau)+\tilde{\mu}_\tau\cdot\sigma_{\mathcal{R}}^{2}(\mtx{X})\underset{\vct{u},\vct{v}\in\mathcal{C}\cap\mathbb{S}^{n-1}}{\sup}\text{ }\vct{u}^T\left(\mtx{I}-\frac{1}{\beta_{s_\tau,m}^2}\mtx{A}_{\mathcal{S}_{\tau}^c}^T\mtx{A}_{\mathcal{S}_{\tau}^c}\right)\vct{v}.
\end{align}

In above, we used the cone-restricted spectral norm of $\mtx{X}$ per Definition \ref{minconeig}. We now focus on simplifying the second term in \ref{mainconv}. To this aim, for a vector $\vct{u}\in\R^n$ define $\vct{u}_{\perp}=\vct{u}-\frac{\vct{u}^T\vct{w}}{\twonorm{\vct{w}}^2}\vct{w}$. We can use this definition to separate the second term in \eqref{mainconv} into two terms as follows
\begin{align*}
\mu_\tau\cdot \tilde{\vct{u}}^T\mtx{X}^T\mtx{A}_{\mathcal{S}_{\tau}^c}^T\mtx{A}_{\mathcal{S}_{\tau}^c}\vct{w}
=&\mu_\tau\twonorm{\mtx{X}\tilde{\vct{u}}} (\vct{u}^T\mtx{A}_{\mathcal{S}_{\tau}^c}^T\mtx{A}_{\mathcal{S}_{\tau}^c}\vct{w})\\
=&\mu_\tau\twonorm{\mtx{X}\tilde{\vct{u}}}(\vct{u}_{\perp}^T\mtx{A}_{\mathcal{S}_{\tau}^c}^T\mtx{A}_{\mathcal{S}_{\tau}^c}\vct{w})+\mu_\tau\frac{\tilde{\vct{u}}^T\mtx{X}^T\vct{w}}{\twonorm{\vct{w}}^2}\twonorm{\mtx{A}_{\mathcal{S}_{\tau}^c}\vct{w}}^2.
%=&\tilde{\mu}_{\tau}\cdot \tilde{\vct{u}}^T\mtx{X}^T\vct{w}-\tilde{\mu}_\tau\cdot\tilde{\vct{u}}^T\mtx{X}^T\left(\mtx{I}-\frac{1}{\beta_{s_\tau,m}^2}\mtx{A}_{\mathcal{S}_{\tau}^c}^T\mtx{A}_{\mathcal{S}_{\tau}^c}\right)\vct{w}\\
%=&\tilde{\mu}_{\tau}\cdot \tilde{\vct{u}}^T\mtx{X}^T\vct{w}-\tilde{\mu}_\tau\cdot\vct{u}^T\left(\mtx{I}-\frac{1}{\beta_{s_\tau,m}^2}\mtx{A}_{\mathcal{S}_{\tau}^c}^T\mtx{A}_{\mathcal{S}_{\tau}^c}\right)\vct{w}
\end{align*}
Thus using $\mu_\tau=\tilde{\mu}_\tau/\beta_{s,m}^2$ we have
\begin{align}
\label{temppf2}
\mu_\tau\cdot \underset{\tilde{\vct{u}}\in-\widetilde{\mathcal{C}}\cap\mathbb{S}^{d-1}}{\sup}\text{ }\tilde{\vct{u}}^T\mtx{X}^T\mtx{A}_{\mathcal{S}_{\tau}^c}^T\mtx{A}_{\mathcal{S}_{\tau}^c}\vct{w}\nonumber\le& \frac{\tilde{\mu}_\tau}{\beta_{s_\tau,m}^2}\cdot\underset{\tilde{\vct{u}}\in-\widetilde{\mathcal{C}}\cap\mathbb{S}^{d-1}}{\sup}\text{ }\twonorm{\mtx{X}\tilde{\vct{u}}}\cdot \underset{\vct{u}\in-\mathcal{C}\cap\mathbb{S}^{n-1}}{\sup}\text{ }\vct{u}_{\perp}^T\mtx{A}_{\mathcal{S}_{\tau}^c}^T\mtx{A}_{\mathcal{S}_{\tau}^c}\vct{w}\nonumber\\
&+\frac{\twonorm{\mtx{A}_{\mathcal{S}_{\tau}^c}\vct{w}}^2}{\beta_{s_\tau,m}^2\cdot\twonorm{\vct{w}}^2}\cdot\tilde{\mu}_\tau\cdot\left(\underset{\tilde{\vct{u}}\in\tilde{\mathcal{C}}\cap\mathbb{S}^{d-1}}{\sup}\text{ }\tilde{\vct{u}}^T\mtx{X}^T\frac{\vct{w}}{\twonorm{\vct{w}}}\right)\twonorm{\vct{w}}\nonumber\\
=&\frac{\tilde{\mu}_\tau\cdot\sigma_{\mathcal{R}}(\mtx{X})}{\beta_{s_\tau,m}^2}\cdot \underset{\vct{u}\in-\mathcal{C}\cap\mathbb{S}^{n-1}}{\sup}\text{ }\vct{u}_{\perp}^T\mtx{A}_{\mathcal{S}_{\tau}^c}^T\mtx{A}_{\mathcal{S}_{\tau}^c}\vct{w}\nonumber\\
&+\frac{1}{\beta_{s_\tau,m}^2}\frac{\twonorm{\mtx{A}_{\mathcal{S}_{\tau}^c}\vct{w}}^2}{\twonorm{\vct{w}}^2}\cdot\tilde{\mu}_{\tau}\cdot\xi(\mtx{X})\twonorm{\vct{w}}.
%=& \tilde{\mu}_{\tau}\cdot\xi(\mtx{X})\twonorm{\vct{w}}+\tilde{\mu}_\tau\cdot\underset{\vct{u}\in\mathcal{C}\cap\mathbb{S}^{d-1}}{\sup}\text{ }\vct{u}^T\left(\mtx{I}-\frac{1}{\beta_{s_\tau,m}^2}\mtx{A}_{\mathcal{S}_{\tau}^c}^T\mtx{A}_{\mathcal{S}_{\tau}^c}\right)\vct{w}
\end{align}
All that remains is to bound the extra additive term in \eqref{temppf1} and the extra additive and multiplicative terms in \eqref{temppf2}. To this aim note that for any $\gamma_\tau$ we have
\begin{align}
\label{intereq}
\vct{u}^T\left(\mtx{I}-\gamma_\tau\mtx{A}_{\mathcal{S}_{\tau}^c}^T\mtx{A}_{\mathcal{S}_{\tau}^c}\right)\vct{v}=&\frac{1}{4}\left(\twonorm{\vct{u}+\vct{v}}^2-\gamma_\tau\twonorm{\mtx{A}_{\mathcal{S}_\tau^c}(\vct{u}+\vct{v})}^2\right)\nonumber\\
&+\frac{1}{4}\left(\gamma_\tau\twonorm{\mtx{A}_{\mathcal{S}_\tau^c}(\vct{u}-\vct{v})}^2-\twonorm{\vct{u}-\vct{v}}^2\right).
\end{align}
To proceed we state a lemma about bounding set-restricted eigenvalues, which is proved in the Appendix section.

\begin{lemma}\label{mainlem}
Let $\mathcal{T}\in\R^n$ and define $\sigma(\mathcal{T}):=\underset{\vct{v}\in\mathcal{T}}{\sup}\text{ }\twonorm{\vct{v}}$. Also assume the random encoding matrix $\mtx{A}\in\R^{m\times n}$ is a matrix with i.i.d.~$\mathcal{N}(0,1)$ entries. Furthermore, define $\alpha_{s,m}=\sqrt{m-2-5s\log\left(\frac{em}{s}\right)}$ and $\beta_{s,m}=\min(\sqrt{3(m-s)\log\left(\frac{em}{(m-s)}\right)},\sqrt{m})$. Then for all $\vct{u}\in\mathcal{T}$ 

\begin{align}
\label{eq1}
\sup_{\mathcal{S}\subset \{1,2,\ldots,m\},\text{ }\abs{\mathcal{S}}= s}\text{ } \twonorm{\mtx{A}_{\mathcal{S}^c}\vct{u}} \le \beta_{s,m}\twonorm{\vct{u}}+\left(\omega(\mathcal{T})+\eta\right),
\end{align}
holds with probability at least $1-2e^{-\frac{\eta^2}{8\sigma^2(\mathcal{T})}}$. Furthermore, for all $\vct{u}\in\mathcal{T}$

\begin{align}
\label{eq2}
\inf_{\mathcal{S}\subset \{1,2,\ldots,m\},\text{ }\abs{\mathcal{S}}= s}\text{ } \twonorm{\mtx{A}_{\mathcal{S}^c}\vct{u}} \ge \alpha_{s,m}\twonorm{\vct{u}}-\left(\omega(\mathcal{T})+\eta\right),
\end{align}
holds with probability at least $1-4e^{-\frac{\eta^2}{8\sigma^2(\mathcal{T})}}$.
\end{lemma}
To use the above lemma, define $\mathcal{T}_{+}= \left(\mathcal{C}\cap\mathbb{S}^{n-1}\right)+\left(\mathcal{C}\cap\mathbb{S}^{n-1}\right)$ and $\mathcal{T}_{-}=\left(\mathcal{C}\cap\mathbb{S}^{n-1}\right)-\left(\mathcal{C}\cap\mathbb{S}^{n-1}\right)$. Also note that $\sigma(\mathcal{T}_{-})\le2$, $\sigma(\mathcal{T}_{+})\le 2$, $\omega(\mathcal{T}_{-})\le 2\omega(\mathcal{C}\cap\mathbb{S}^{n-1})$, $\omega(\mathcal{T}_{+})\le 2\omega(\mathcal{C}\cap\mathbb{S}^{n-1})$, and $\vct{u}+\vct{v}\in\mathcal{T}_{+}$. Thus by Lemma \ref{mainlem} equation \eqref{eq2}
\begin{align}
\label{eqconc1}
\twonorm{\vct{u}+\vct{v}}^2-\gamma_\tau\twonorm{\mtx{A}_{\mathcal{S}_\tau^c}(\vct{u}+\vct{v})}^2\nonumber\le&\twonorm{\vct{u}+\vct{v}}^2-\gamma_\tau\left(\alpha_{s_\tau,m}\twonorm{\vct{u}+\vct{v}}-\left(\omega(\mathcal{T}_{+})+\eta\right)\right)^2\nonumber\\
\le&\left(1-\gamma_\tau\alpha_{s_\tau,m}^2\right)\twonorm{\vct{u}+\vct{v}}^2\nonumber\\
&+2\gamma_\tau\alpha_{s_\tau,m}\left(\omega(\mathcal{T}_{+})+\eta\right)\twonorm{\vct{u}+\vct{v}}-\gamma_\tau\left(\omega(\mathcal{T}_{+})+\eta\right)^2\nonumber\\
=&\left(1-\gamma_\tau\alpha_{s_\tau,m}^2\right)\twonorm{\vct{u}+\vct{v}}^2\nonumber\\
&+2\gamma_\tau\alpha_{s_\tau,m}\left(2\omega(\mathcal{C}\cap\mathbb{S}^{n-1})+\eta\right)\twonorm{\vct{u}+\vct{v}}\nonumber\\
&-\gamma_\tau\left(2\omega(\mathcal{C}\cap\mathbb{S}^{n-1})+\eta\right)^2.
\end{align}
holds with probability at least $1-4e^{-\frac{\eta^2}{32}}$. Also, $\vct{u}-\vct{v}\in\mathcal{T}_{-}$, thus by Lemma \ref{mainlem} equation \eqref{eq1}
\begin{align}
\label{eqconc2}
\gamma_\tau\twonorm{\mtx{A}_{\mathcal{S}_\tau^c}(\vct{u}-\vct{v})}^2-\twonorm{\vct{u}-\vct{v}}^2\nonumber
\le&\left(\gamma_\tau\beta_{s_\tau,m}^2-1\right)\twonorm{\vct{u}-\vct{v}}^2\nonumber+2\gamma_\tau\beta_{s_\tau,m}\left(\omega(\mathcal{T}_{-})+\eta\right)\twonorm{\vct{u}-\vct{v}}\nonumber\\
&+\gamma_\tau\left(\omega(\mathcal{T}_{-})+\eta\right)^2\nonumber\\
=&\left(\gamma_\tau\beta_{s_\tau,m}^2-1\right)\twonorm{\vct{u}-\vct{v}}^2\nonumber+2\gamma_\tau\beta_{s_\tau,m}\left(2\omega(\mathcal{C}\cap\mathbb{S}^{n-1})+\eta\right)\twonorm{\vct{u}-\vct{v}}\nonumber\\
&+\gamma_\tau\left(2\omega(\mathcal{C}\cap\mathbb{S}^{n-1})+\eta\right)^2
\end{align}
holds with probability at least $1-2e^{-\frac{\eta^2}{32}}$. Plugging these bounds into \eqref{intereq} with $\twonorm{\vct{u}+\vct{v}}\le 2$ and $\twonorm{\vct{u}-\vct{v}}\le 2$ and using the short-hand $\omega:=\omega(\mathcal{C}\cap\mathbb{S}^{n-1})$ we conclude that for $\gamma_\tau=\frac{1}{\beta_{s_\tau,m}^2}$
\begin{align}
\vct{u}^T(\mtx{I}-\gamma_\tau\mtx{A}_{\mathcal{S}_{\tau}^c}^T\mtx{A}_{\mathcal{S}_{\tau}^c})\vct{v}\nonumber\le& \frac{1}{4}\left(1-\gamma_\tau\alpha_{s_\tau,m}^2\right)\twonorm{\vct{u}+\vct{v}}^2\nonumber
+\frac{1}{2}\gamma_\tau\alpha_{s_\tau,m}\left(2\omega+\eta\right)\twonorm{\vct{u}+\vct{v}}\nonumber\\
&+\frac{1}{4}\left(\gamma_\tau\beta_{s_\tau,m}^2-1\right)\twonorm{\vct{u}-\vct{v}}^2\nonumber+\frac{1}{2}\gamma_\tau\beta_{s_\tau,m}\left(2\omega+\eta\right)\twonorm{\vct{u}-\vct{v}}\nonumber\\
\le& 1-\frac{\alpha_{s,m}^2}{\beta_{s_\tau,m}^2}+4\frac{\left(\omega+\frac{\eta}{2}\right)}{\beta_{s_\tau,m}}
%\le&1-\frac{\alpha_{s_\tau,m}^2}{\beta_{s_\tau,m}^2}
\end{align}
holds with probability at least $1-6e^{-\frac{\eta^2}{32}}$. Using a change of variable $\eta$ to $2\eta$ together with the fact that $\sqrt{m-s_\tau}\le\beta_{s_\tau,m}\le \sqrt{m}$ we arrive at
\begin{align}
\vct{u}^T(\mtx{I}-\mu_\tau\mtx{A}_{\mathcal{S}_{\tau}^c}^T\mtx{A}_{\mathcal{S}_{\tau}^c})\vct{v}\nonumber&\le\frac{2+5s_\tau\log(em/s_\tau)}{m}+4\sqrt{\frac{m_0}{m-s_\tau}},
\end{align}
holds with probability at least $1-6e^{-\frac{\eta^2}{8}}$ completing the proof of the bound on the extra term of \eqref{temppf1}. 

Now we focus on the extra additive and multiplicative terms in \eqref{temppf2}. We begin with the additive term. To this aim note that since $\vct{u}_{\perp}$ is orthagonal to $\vct{w}$, $\vct{u}_{\perp}^T\mtx{A}_{\mathcal{S}_{\tau}^c}^T\mtx{A}_{\mathcal{S}_{\tau}^c}\vct{w}$ has the same distribution as $\twonorm{\vct{w}}\vct{u}_{\perp}^T\mtx{A}_{\mathcal{S}_{\tau}^c}^T\vct{a}$ with $\vct{a}$ distributed as $\vct{a}\sim\mathcal{N}(0,\mtx{I}_{m-\abs{\mathcal{S}_\tau}})$ and independent from $\mtx{A}$. Similarly, $\twonorm{\vct{w}}\vct{u}_{\perp}^T\mtx{A}_{\mathcal{S}_{\tau}^c}^T\vct{a}$ has the same distribution as $\twonorm{\vct{a}}\twonorm{\vct{w}}(\vct{u}_{\perp}^T\vct{g})$ with $\vct{g}\sim\mathcal{N}(\vct{0},\mtx{I}_n)$. Therefore,
\begin{align}
\label{subtemp1}
\underset{\vct{u}\in-\mathcal{C}\cap\mathbb{S}^{n-1}}{\sup}\text{ }\vct{u}_{\perp}^T\mtx{A}_{\mathcal{S}_{\tau}^c}^T\mtx{A}_{\mathcal{S}_{\tau}^c}\vct{w}\nonumber=&\twonorm{\vct{a}}\twonorm{\vct{w}}\cdot \left( \underset{\vct{u}\in-\mathcal{C}\cap\mathbb{S}^{n-1}}{\sup}\text{ }\vct{u}_{\perp}^T\vct{g}\right)\nonumber\\
\le& \twonorm{\vct{a}}\twonorm{\vct{w}}\cdot \left(\omega+\eta\right)\nonumber\\
\le& \sqrt{2(m-\abs{\mathcal{S}}_{\tau})}\twonorm{\vct{w}}\cdot \left(\omega+\eta\right)
\end{align}
holds with probability at least $1-e^{-\frac{\eta^2}{2}}-e^{-\frac{m}{2}}$.

We now focus on the extra multiplicative term in \eqref{temppf2}. To this aim note that since $\vct{w}$ is fixed $\mtx{A}_{\mathcal{S}_{\tau}^c}\vct{w}$ is distributed as $\twonorm{\vct{w}}\vct{a}$ with $\vct{a}$ distributed as $\mathcal{N}(0,\mtx{I}_{m-\abs{\mathcal{S}_\tau}})$. Therefore,
\begin{align}
\label{subtemp2}
\frac{\twonorm{\mtx{A}_{\mathcal{S}_{\tau}^c}\vct{w}}^2}{\twonorm{\vct{w}}^2}=\twonorm{\vct{a}}^2\le 2(m-\abs{\mathcal{S}_\tau}),
\end{align}
holds with probability at least $1-e^{-\frac{m}{2}}$. Plugging \eqref{subtemp1} and \eqref{subtemp2} into \eqref{temppf2} we conclude that 
\begin{align}
\mu_\tau\cdot \underset{\tilde{\vct{u}}\in-\widetilde{\mathcal{C}}\cap\mathbb{S}^{d-1}}{\sup}\text{ }\tilde{\vct{u}}^T\mtx{X}^T\mtx{A}_{\mathcal{S}_{\tau}^c}^T\mtx{A}_{\mathcal{S}_{\tau}^c}\vct{w}\nonumber\le&\sqrt{2}\frac{\tilde{\mu}_\tau\cdot\sigma_{\mathcal{R}}(\mtx{X})}{\beta_{s_\tau,m}^2}\cdot \sqrt{m-\abs{\mathcal{S}_\tau}}\twonorm{\vct{w}}\cdot \left(\omega+\eta\right)\nonumber\\&+\frac{2(m-\abs{\mathcal{S}_\tau})}{\beta_{s_\tau,m}^2}\cdot\tilde{\mu}_{\tau}\cdot\xi(\mtx{X})\twonorm{\vct{w}}\nonumber\\
\le&\tilde{\mu}_\tau\cdot\sigma_{\mathcal{R}}(\mtx{X})\sqrt{\frac{m_0}{2(m-s_\tau)\log^2\left(\frac{em}{m-s_\tau}\right)}}\twonorm{\vct{w}}\nonumber\\&+\frac{1}{\log\left(\frac{em}{m-s_\tau}\right)}\tilde{\mu}_{\tau}\cdot\xi(\mtx{X})\twonorm{\vct{w}}\nonumber\\
\le&\tilde{\mu}_{\tau}\cdot\xi(\mtx{X})\twonorm{\vct{w}}+\frac{1}{\sqrt{2}}\tilde{\mu}\cdot\sigma_{\mathcal{R}}(\mtx{X})\sqrt{\frac{m_0}{m-s_\tau}},
\end{align}
holds with probability at least $1-e^{-\frac{\eta^2}{2}}-e^{-\frac{m}{2}}$.

\newpage

\section{Acknowledgements}
M. Soltanolkotabi is supported by the Packard Fellowship in Science and Engineering, a Sloan Research Fellowship in Mathematics, an NSF-CAREER under award \#1846369, the Air Force Office of Scientific Research Young Investigator Program (AFOSR-YIP)
under award \#FA9550-18-1-0078, an NSF-CIF award \#1813877, and a Google faculty research award. 

\bibliography{Bibfiles-2}
\bibliographystyle{plain}
\section{Appendix}

%\label{subsecpf}
In this section we aim to prove Lemma \ref{mainlem} stated in the proofs section.
%. We restate this lemma for the convenience of the readers.
%\begin{lemma}\label{mainlem1}
%Let $\mathcal{T}\in\R^n$ and define $\sigma(\mathcal{T}):=\underset{\vct{v}\in\mathcal{T}}{\sup}\text{ }\twonorm{\vct{v}}$. Also assume $\mtx{A}\in\R^{m\times n}$ is a matrix with i.i.d.~$\mathcal{N}(0,1)$ entries. Furthermore, define $\alpha_{s,m}=\sqrt{m-2-5s\log\left(\frac{em}{s}\right)}$ and \\ $\beta_{s,m}=\min\left(\sqrt{3(m-s)\log\left(\frac{em}{(m-s)}\right)},\sqrt{m}\right)$. Then for all $\vct{u}\in\mathcal{T}$ 
%\vspace{0.2cm}
%\begin{align}
%\label{eq1}
%\sup_{\mathcal{S}\subset \{1,2,\ldots,m\},\text{ }\abs{\mathcal{S}}= s}\text{ } \twonorm{\mtx{A}_{\mathcal{S}^c}\vct{u}} \le \beta_{s,m}\twonorm{\vct{u}}+\left(\omega(\mathcal{T})+\eta\right),
%\end{align}
%holds with probability at least $1-2e^{-\frac{\eta^2}{8\sigma^2(\mathcal{T})}}$. Furthermore, for all $\vct{u}\in\mathcal{T}$
%\vspace{0.2cm}
%\begin{align}
%\label{eq2}
%\inf_{\mathcal{S}\subset \{1,2,\ldots,m\},\text{ }\abs{\mathcal{S}}= s}\text{ } \twonorm{\mtx{A}_{\mathcal{S}^c}\vct{u}} \ge \alpha_{s,m}\twonorm{\vct{u}}-\left(\omega(\mathcal{T})+\eta\right),
%\end{align}
%holds with probability at least $1-4e^{-\frac{\eta^2}{8\sigma^2(\mathcal{T})}}$.
%\end{lemma}
Our proof is related to the proof of Gordon's celebrated escape through the mesh \cite{Gor}[Theorem A]. We will first show the bound \eqref{eq1}. To this aim we make use of Slepian's lemma stated below.
\begin{lemma}[Slepian's inequality] \cite{Banach1}[Section 3.3]  If $X_t$ and $Y_t$ are a.s. bounded, Gaussian processes on $T$ such that $\E[X_t]=\E[Y_t]$ and $\E[X_t^2]=\E[Y_t^2]$ for all $t\in T$ and
\begin{align*}
\E[(X_t-X_s)^2]\le\E[(Y_t-Y_s)^2],
\end{align*}
for all $s,t\in T$, then for all real $t$,
\begin{align}
\label{firstSlep}
\E[\underset{t\in T}{\sup}\text{ }X_t]\le\E[\underset{t\in T}{\sup}\text{ }Y_t].
\end{align}
Furthermore,
\begin{align}
\label{secondSlep}
%\mathbb{P}\Big\{\underset{t\in T}{\sup}\text{ }X_t> \eta\Big\}\le\mathbb{P}\Big\{\underset{t\in T}{\sup}\text{ }Y_t> \eta\Big\}.
\mathbb{P}\bigg\{\underset{t\in T}{\bigcup}[X_t>\eta_t]\bigg\}\le\mathbb{P}\bigg\{\underset{t\in T}{\bigcup}[Y_t>\eta_t]\bigg\}.
\end{align}
\end{lemma}
Define $\mtx{I}_{\mathcal{S}^c}\in\R^{(m-s)\times n}$ as the part of the identity matrix that keeps the rows indexed by $\mathcal{S}^c$. For $\vct{u}\in\mathcal{T}$ and $\vct{v}\in\mathbb{S}^{m-s-1}=\{\vct{v}\in\R^{m-s};\text{ }\twonorm{\vct{v}}=1\}$, we define three Gaussian processes
\begin{align*}
X_{(\vct{u},\mathcal{S}),\vct{v}}=\vct{v}^*\mtx{I}_{\mathcal{S}^c}\mtx{A}\vct{u},\quad Y_{(\vct{u},\mathcal{S}),\vct{v}}=\twonorm{\vct{u}}\vct{v}^*\mtx{I}_{\mathcal{S}^c}\vct{a}+\vct{g}^*\vct{u}\quad\text{and}\quad Z_{(\vct{u},\mathcal{S}),\vct{v}}=\twonorm{\vct{u}}(\vct{v}^*\mtx{I}_{\mathcal{S}^c}\vct{a}-\beta_{s,m})+\vct{g}^*\vct{u}.
\end{align*}
Here $\vct{a}\in\R^m$ is distributed as $\mathcal{N}(\vct{0},\mtx{I}_m)$ and $\vct{g}\in\R^n$ is distributed as $\mathcal{N}(\vct{0},\mtx{I}_n)$. It follows that for all $\vct{u},\tilde{\vct{u}}\in\mathcal{T}$, $\vct{v},\tilde{\vct{v}}\in\mathbb{S}^{m-s-1}$ and $\mathcal{S}, \tilde{\mathcal{S}}\subset\{1,2,\ldots,m\}$, we have
\begin{align}
\label{SlepH}
\mathbb{E}\abs{Y_{(\vct{u},\mathcal{S}),\vct{v}}-Y_{(\tilde{\vct{u}},\tilde{\mathcal{S}}),\tilde{\vct{v}}}}^2-\mathbb{E}\abs{X_{(\vct{u},\mathcal{S}),\vct{v}}-X_{(\tilde{\vct{u}},\tilde{\mathcal{S}}),\tilde{\vct{v}}}}^2=&\twonorm{\twonorm{\vct{u}}\mtx{I}_{\mathcal{S}^c}^T\vct{v}-\twonorm{\tilde{\vct{u}}}\mtx{I}_{\tilde{\mathcal{S}}^c}^T\tilde{\vct{v}}}^2+\twonorm{\vct{u}-\tilde{\vct{u}}}^2\nonumber\\
&-\fronorm{\vct{u}\left(\mtx{I}_{\mathcal{S}^c}^T\vct{v}\right)^T-\tilde{\vct{u}}\left(\mtx{I}_{\tilde{\mathcal{S}}^c}^T\tilde{\vct{v}}\right)^T}^2\nonumber\\
=&\left(\twonorm{\vct{u}}^2+\twonorm{\tilde{\vct{u}}}^2\right)-2\twonorm{\vct{u}}\twonorm{\tilde{\vct{u}}}\langle \mtx{I}_{\mathcal{S}^c}^T\vct{v}, \mtx{I}_{\tilde{\mathcal{S}}^c}^T\tilde{\vct{v}}\rangle-2\langle\vct{u},\tilde{\vct{u}}\rangle\nonumber\\
&+2\langle\vct{u},\tilde{\vct{u}}\rangle\langle \mtx{I}_{\mathcal{S}^c}^T\vct{v}, \mtx{I}_{\tilde{\mathcal{S}}^c}^T\tilde{\vct{v}}\rangle\nonumber\\
=&\left(\twonorm{\vct{u}}^2+\twonorm{\tilde{\vct{u}}}^2\right)-2\twonorm{\vct{u}}\twonorm{\tilde{\vct{u}}}\langle \mtx{I}_{\mathcal{S}^c}^T\vct{v}, \mtx{I}_{\tilde{\mathcal{S}}^c}^T\tilde{\vct{v}}\rangle\nonumber\\
&-2\langle\vct{u},\tilde{\vct{u}}\rangle\left(1-\langle \mtx{I}_{\mathcal{S}^c}^T\vct{v}, \mtx{I}_{\tilde{\mathcal{S}}^c}^T\tilde{\vct{v}}\rangle\right)\nonumber\\
\ge&2\twonorm{\vct{u}}\twonorm{\tilde{\vct{u}}}-2\twonorm{\vct{u}}\twonorm{\tilde{\vct{u}}}\langle \mtx{I}_{\mathcal{S}^c}^T\vct{v}, \mtx{I}_{\tilde{\mathcal{S}}^c}^T\tilde{\vct{v}}\rangle\nonumber\\
&-2\langle\vct{u},\tilde{\vct{u}}\rangle\left(1-\langle \mtx{I}_{\mathcal{S}^c}^T\vct{v}, \mtx{I}_{\tilde{\mathcal{S}}^c}^T\tilde{\vct{v}}\rangle\right)\nonumber\\
=& 2\left(\twonorm{\vct{u}}\twonorm{\tilde{\vct{u}}}-\langle\vct{u},\tilde{\vct{u}}\rangle\right)\left(1-\langle \mtx{I}_{\mathcal{S}^c}^T\vct{v}, \mtx{I}_{\tilde{\mathcal{S}}^c}^T\tilde{\vct{v}}\rangle\right)\nonumber\\
\ge& 0.
\end{align}
\\
AIt is trivial to check that $\E[X_{(\vct{u},\mathcal{S}),\vct{v}}]=\E[Y_{(\vct{u},\mathcal{S}),\vct{v}}]$ and $\E[X_{(\vct{u},\mathcal{S}),\vct{v}}^{2}]=\E[Y_{(\vct{u},\mathcal{S}),\vct{v}}^{2}]$ for all $\vct{u}\in\mathcal{T},\vct{v}\in\mathbb{S}^{m-s-1}$ and
$\mathcal{S}\subset\{1,2,\ldots,m\}$. Thus, the two Gaussian processes $X_{(\vct{u},\mathcal{S}),\vct{v}}$ and $Y_{(\vct{u},\mathcal{S}),\vct{v}}$ obey the three assumptions of Slepian's inequality.

%Now by utilizing Dudley's inequality and the definition of $\beta_{s,m}$ we have 

%%Hence $f(\vct{a})$ is a Lipschitz function of a Gaussian random variable. Thus, the random variable $Z:=f(\vct{a})$ obeys
%%\begin{align*}
%%\text{Var}(Z)\le 1\quad\Rightarrow\quad \E[Z^2]-\left(\E[Z]\right)^2\le1\quad\Rightarrow\quad \E[Z]\ge\sqrt{\E[Z^2]-1}.
%%\end{align*}
%%Thus
%begin{align}
%\label{eqin1}
%\E\bigg[\underset{\mathcal{S}\subset \{1,2,\ldots,m\},\text{ }\abs{\mathcal{S}}=s}{\sup}\text{ }\twonorm{\vct{a}_{\mathcal{S}^c}}\bigg]\le&\sqrt{\E\bigg[\underset{\mathcal{S}\subset \{1,2,\ldots,m\},\text{ }\abs{\mathcal{S}}=s}{\sup}\text{ }\twonorm{\vct{a}_{\mathcal{S}^c}}^2\bigg]},\nonumber\\
%\le&\sqrt{4(m-s)\log\left(\frac{em}{(m-s)}\right)},\nonumber\\
%\le&\beta_{s,m}.
%\end{align}

Now define the function
$f(\vct{x})=\underset{\mathcal{S}\subset \{1,2,\ldots,m\},\text{ }\abs{\mathcal{S}}=s}{\sup}\text{ }\twonorm{\vct{x}_{\mathcal{S}^c}}$ and let 

\begin{align*}
\mathcal{S}_{\vct{x}}^c=\underset{\mathcal{S}\subset \{1,2,\ldots,m\},\text{ }\abs{\mathcal{S}}=s}{\arg\max} f(\vct{x})\quad\text{and}\quad\mathcal{S}_{\vct{y}}^c=&\underset{\mathcal{S}\subset \{1,2,\ldots,m\},\text{ }\abs{\mathcal{S}}=s}{\arg\max} f(\vct{y}).
\end{align*}

%In order to show that $f$ is $1$-Lipschitz, without loss of generality assume $f(\vct{x})\ge f(\vct{y})$. Thus,

%\begin{align*}
%\abs{f(\vct{x})-f(\vct{y})}=f(\vct{x})-f(\vct{y})=\twonorm{\vct{x}_{\mathcal{S}_{\vct{x}}^c}}-\twonorm{\vct{y}_{\mathcal{S}_{\vct{y}}^c}}\le \twonorm{\vct{x}_{\mathcal{S}_{\vct{x}}^c}}-\twonorm{\vct{y}_{\mathcal{S}_{\vct{x}}^c}}\le \twonorm{\left(\vct{x}-\vct{y}\right)_{\mathcal{S}_{\vct{x}}^c}}\le\twonorm{\vct{x}-\vct{y}}.
%\end{align*}

%e note that since $f$ is $1$-Lipschitz by standard concentration of measure for Gaussian random variables
%By concentration of Lipschitz functions of Gaussians

%\begin{align}
%\mathbb{P}\Big\{f(\vct{a})\ge \E[f(\vct{a})]+\eta\Big\}\le e^{-\frac{\eta^2}{2}}.
%\end{align}

We wish to bound $f(\vct{a})$ with high probability. To this aim first note that by concentration of Lipschitz functions of Gaussians

\begin{align*}
\mathbb{P}\Big\{\twonorm{\vct{a}_{\mathcal{S}^c}}-\E[\twonorm{\vct{a}_{\mathcal{S}^c}}\ge\delta\Big\}\le e^{\frac{-\delta^2}{2}}
\end{align*}

Note that since $\E[\twonorm{\vct{a}_{\mathcal{S}^c}}]\le\sqrt{\E[\twonorm{\vct{a}_{\mathcal{S}^c}}^2]}=\sqrt{m-s}$, by substituting $\delta=\eta+\sqrt{2(m-s)\log\left(\frac{em}{m-s}\right)}$ we have

\begin{align*}
\mathbb{P}\Bigg\{\twonorm{\vct{a}_{\mathcal{S}^c}}-\sqrt{m-s}\ge\eta+\sqrt{2(m-s)\log\left(\frac{em}{m-s}\right)}\Bigg\}\le& e^{\frac{-\left(\eta+\sqrt{2(m-s)\log\left(\frac{em}{m-s}\right)}\right)^{2}}{2}}\nonumber\\\le&e^{\frac{-\left(\sqrt{2(m-s)\log\left(\frac{em}{m-ms}\right)}\right)^2}{2}}e^{\frac{-\eta^2}{2}}
\end{align*}

Using union bound, we have

\begin{align*}
\mathbb{P}\Bigg\{\underset{\mathcal{S}\subset \{1,2,\ldots,m\},\text{ }\abs{\mathcal{S}}=s}{\sup}\text{ }\twonorm{\vct{a}_{\mathcal{S}^c}}\ge\eta+\sqrt{m-s}+\sqrt{2(m-s)\log\left(\frac{em}{m-s}\right)}\Bigg\}\le& {m \choose m-s} e^{\frac{-\left(\sqrt{2(m-s)\log\left(\frac{em}{m-s}\right)}\right)^2}{2}}e^{\frac{-\eta^2}{2}}\nonumber\\=&{m \choose m-s}{\left(\frac{em}{m-s}\right)}^{-s}e^{\frac{-\eta^2}{2}}\nonumber\\
\le&e^{\frac{-\eta^2}{2}}
\end{align*}

We thus conclude that
\begin{align}
\label{mmmtemp}
\mathbb{P}\Bigg\{\underset{\mathcal{S}\subset \{1,2,\ldots,m\},\text{ }\abs{\mathcal{S}^c}=s}{\sup}\text{ }\twonorm{\vct{a}_{\mathcal{S}^c}}\ge\eta+\sqrt{m-s}+\sqrt{2(m-s)\log\left(\frac{em}{m-s}\right)}\Bigg\}\le&e^{\frac{-\eta^2}{2}}.
\end{align}
Also note that
\begin{align*}
\sqrt{m-s}+\sqrt{2(m-s)\log\left(\frac{em}{m-s}\right)} \ \le \text{min}\left(\sqrt{3(m-s)\log\left(\frac{em}{m-s}\right)},m\right):=\beta_{s,m}
\end{align*}
The latter together with \eqref{mmmtemp} allows us to conclude that
\begin{align}
\label{msmst}
\mathbb{P}\Bigg\{f(\vct{a})\ge \beta_{s,m}+\eta\Bigg\}\le e^{-\frac{\eta^2}{2}}.
\end{align}

Now consider the relationship of following sets

\begin{align*}
\Big\{\vct{a}: \quad\twonorm{\vct{u}}f(\vct{a})\ge \twonorm{\vct{u}}\beta_{s,m}+\eta\Big\}\subset&\Big\{\vct{a}: \quad\twonorm{\vct{u}}f(\vct{a})\ge \twonorm{\vct{u}}\beta_{s,m}+\twonorm{\vct{u}}\frac{\eta}{\sigma(\mathcal{T})}\Big\},\\
=&\Big\{\vct{a}: \quad f(\vct{a})\ge \beta_{s,m}+\frac{\eta}{\sigma(\mathcal{T})}\Big\}.\\
%=&\Big\{\vct{a}: \quad f(\vct{a})\ge \E[f(\vct{a})]+\frac{\eta}{\sigma(\mathcal{T})}\Big\}.
\end{align*}

Furthermore, note for every $\vct{u}\in\mathcal{T}$, $\Big\{\vct{a}: \quad\twonorm{\vct{u}}f(\vct{a})\ge \twonorm{\vct{u}}\beta_{s,m}+\eta\Big\}$ is a subset of $\Big\{\vct{a}: \quad f(\vct{a})\ge \beta_{s,m}+\frac{\eta}{\sigma(\mathcal{T})}\Big\}$. Combining the latter with \eqref{msmst} we arrive at

\begin{align*}
\mathbb{P}\Bigg\{\underset{\vct{u}\in\mathcal{T}}{\bigcup}\big\{\vct{a}:\quad f(\vct{a})\twonorm{\vct{u}}>\twonorm{\vct{u}} \beta_{s,m}+\eta_1\big\}\Bigg\}\le\mathbb{P}\Bigg\{\vct{a}:\text{ }f(\vct{a})\ge \beta_{s,m}+\frac{\eta_1}{\sigma(\mathcal{T})}\Bigg\}\le e^{-\frac{\eta_1^2}{2\sigma^2(\mathcal{T})}},
\end{align*}

which immediately implies
\begin{align}
\label{myfirstineqGT}
\mathbb{P}\Big\{\underset{\vct{u}\in\mathcal{T}}{\max}\text{ }\twonorm{\vct{u}}\left(f(\vct{a})-\beta_{s,m}\right)>\frac{\eta}{2}\Big\}\le e^{-\frac{\eta^2}{8\sigma^2(\mathcal{T})}}.
\end{align}

Also by the concentration of Lipschitz functions of Gaussians for the function $\underset{\vct{u}\in\mathcal{T}}{\max} \text{ }\left(\vct{g}^*\vct{u}\right)$ and the definition of Gaussian width, we have

\begin{align}
\label{mysecondineqGT}
\mathbb{P}\Big\{\text{ }\underset{\vct{u}\in\mathcal{T}}{\max} \text{ }\left(\vct{g}^*\vct{u}\right)>\omega(\mathcal{T})+\frac{\eta}{2}\Big\}=\mathbb{P}\Big\{\text{ }\underset{\vct{u}\in\mathcal{T}}{\max} \text{ }\left(\vct{g}^*\vct{u}\right)>\E\big[\underset{\vct{u}\in\mathcal{T}}{\max} \text{ }\left(\vct{g}^*\vct{u}\right)\big]+\frac{\eta}{2}\Big\}\le e^{-\frac{\eta^2}{8\sigma^2(\mathcal{T})}}.
\end{align}

So far, we have obtained upper bounds on the probability of sets $\Big\{\underset{\vct{u}\in\mathcal{T}}{\max}\text{ }\twonorm{\vct{u}}\left(f(\vct{a})-\beta_{s,m}\right)>\frac{\eta}{2}\Big\}$ and $\ \Big\{\text{ }\underset{\vct{u}\in\mathcal{T}}{\max} \text{ }\left(\vct{g}^*\vct{u}\right)>\omega(\mathcal{T})+\frac{\eta}{2}\Big\}$. In order to utilize these two sets we combine them in the following way. Note that if

\begin{align*}
\underset{\vct{u}\in\mathcal{T}}{\max}\text{ }\left(\twonorm{\vct{u}}\left(f(\vct{a})-\beta_{s,m}\right)+\vct{g}^*\vct{u}\right)>\eta+\omega(\mathcal{T}),
\end{align*}

then we have either $\ \underset{\vct{u}\in\mathcal{T}}{\max}\text{ }\twonorm{\vct{u}}\left(f(\vct{a})-\beta_{s,m}\right)>\frac{\eta}{2}$ \ or \ $\underset{\vct{u}\in\mathcal{T}}{\max} \text{ }\left(\vct{g}^*\vct{u}\right)>\omega(\mathcal{T})+\frac{\eta}{2}$, which implies that

\begin{align*}
\big\{\vct{a},\vct{g}:\text{ }\underset{\vct{u}\in\mathcal{T}}{\max}\text{ }\left(\twonorm{\vct{u}}f(\vct{a})+\vct{g}^*\vct{u}-\beta_{s,m}\twonorm{\vct{u}}\right)>\omega(\mathcal{T})+\eta\big\}
\end{align*}
is a subset of 

\begin{align*}
\big\{\vct{a},\vct{g}:\text{ }\underset{\vct{u}\in\mathcal{T}}{\max}\text{ }\twonorm{\vct{u}}\left(f(\vct{a})-\beta_{s,m}\right)>\frac{\eta}{2}\big\}\bigcup\big\{\vct{a},\vct{g}:\text{ }\underset{\vct{u}\in\mathcal{T}}{\max} \text{ }\left(\vct{g}^*\vct{u}\right)>\omega(\mathcal{T})+\frac{\eta}{2}\big\}.
\end{align*}

Using the latter together with \eqref{myfirstineqGT} and \eqref{mysecondineqGT} and using the independence of $\vct{a}$ and $\vct{g}$ we have

\begin{align*}
\mathbb{P}\Big\{\underset{\vct{u}\in\mathcal{T}}{\max}\text{ }\left(\twonorm{\vct{u}}\left(f(\vct{a})-\beta_{s,m}\right)+\vct{g}^*\vct{u}\right)>\omega(\mathcal{T})+\eta\Big\}\le& \mathbb{P}\Big\{\underset{\vct{u}\in\mathcal{T}}{\max}\text{ }\twonorm{\vct{u}}\left(f(\vct{a})-\beta_{s,m}\right)>\frac{\eta}{2}\Big\}\\&+\mathbb{P}\Big\{\text{ }\underset{\vct{u}\in\mathcal{T}}{\max} \text{ }\left(\vct{g}^*\vct{u}\right)>\omega(\mathcal{T})+\frac{\eta}{2}\Big\}\nonumber\\
\le& 2e^{-\frac{\eta^2}{8\sigma^2(\mathcal{T})}}.
\end{align*}

Using the definition of $f(\vct{x})=\underset{\mathcal{S}\subset \{1,2,\ldots,m\},\text{ }\abs{\mathcal{S}}=s}{\sup}\text{ }\twonorm{\vct{x}_{\mathcal{S}^c}}$, the latter statement can be rewritten in the form

\begin{align}
\label{mythirdineqGT}
\mathbb{P}\Bigg\{\underset{\mathcal{S}\subset \{1,2,\ldots,m\},\text{ }\abs{\mathcal{S}}=s}{\sup}\text{ }\underset{\vct{u}\in\mathcal{T}}{\max}\text{ }\twonorm{\vct{u}}\left(\twonorm{\vct{a}_{\mathcal{S}^c}}-\beta_{s,m}\right)+\vct{g}^*\vct{u}>\omega(\mathcal{T})+\eta\Bigg\}\le2e^{-\frac{\eta^2}{8\sigma^2(\mathcal{T})}}.
\end{align}

As we defined the Gaussian process $Z_{(\vct{u},\mathcal{S}),\vct{v}}=\twonorm{\vct{u}}(\vct{v}^*\mtx{I}_{\mathcal{S}^c}\vct{a}-\beta_{s,m})+\vct{g}^*\vct{u}$, we can write

\begin{align*}
\underset{\mathcal{S}\subset \{1,2,\ldots,m\},\text{ }\abs{\mathcal{S}}=s}{\sup}\text{ }\underset{\vct{u}\in\mathcal{T},\text{ }\vct{v}\in\mathbb{S}^{m-s-1}}{\max}\text{ }Z_{(\vct{u},\mathcal{S}),\vct{v}}=\underset{\mathcal{S}\subset \{1,2,\ldots,m\},\text{ }\abs{\mathcal{S}}=s}{\sup}\text{ }\underset{\vct{u}\in\mathcal{T}}{\max}\text{ }\twonorm{\vct{u}}\left(\twonorm{\vct{a}_{\mathcal{S}^c}}-\beta_{s,m}\right)+\vct{g}^*\vct{u}.
\end{align*}

This together with \eqref{mythirdineqGT} implies
\begin{align*}
\mathbb{P}\Bigg\{\underset{\mathcal{S}\subset \{1,2,\ldots,m\},\text{ }\abs{\mathcal{S}}=s}{\sup}\text{ }\underset{\vct{u}\in\mathcal{T},\text{ }\vct{v}\in\mathbb{S}^{m-s-1}}{\max}&\text{ }Z_{(\vct{u},\mathcal{S}),\vct{v}}>\omega(\mathcal{T})+\eta\Bigg\}\le 2e^{-\frac{\eta^2}{8\sigma^2(\mathcal{T})}}.
\end{align*}
%&\Rightarrow\quad

Also, $Z_{(\vct{u},\mathcal{S}),\vct{v}}=Y_{(\vct{u},\mathcal{S}),\vct{v}}-\beta_{s,m}\twonorm{\vct{u}}$ implies that
\begin{align*}
\mathbb{P}\Bigg\{\underset{\mathcal{S}\subset \{1,2,\ldots,m\},\text{ }\abs{\mathcal{S}}=s,\text{ }\vct{u}\in\mathcal{T},\vct{v}\in\mathbb{S}^{m-s-1}}{\bigcup}[Y_{(\vct{u},\mathcal{S}),\vct{v}}>\beta_{s,m}\twonorm{\vct{u}}+\omega(\mathcal{T})+\eta]\Bigg\}\le 2e^{-\frac{\eta^2}{8\sigma^2(\mathcal{T})}}.
\end{align*}

As we noted that the two Gaussian processes $X_{(\vct{u},\mathcal{S}),\vct{v}}$ and $Y_{(\vct{u},\mathcal{S}),\vct{v}}$ have the three assumptions of Slepian's inequality, we can use Slepian's second inequality \eqref{secondSlep} with $\eta_{\vct{u},\vct{v}}=\beta_{s,m}\twonorm{\vct{u}}+\eta+\omega(\mathcal{T})$. This implies that
\begin{align*}
\mathbb{P}\Bigg\{\underset{\mathcal{S}\subset \{1,2,\ldots,m\},\text{ }\abs{\mathcal{S}}=s,\text{ }\vct{u}\in\mathcal{T},\vct{v}\in\mathbb{S}^{m-s-1}}{\bigcup}&[X_{(\vct{u},\mathcal{S}),\vct{v}}>\beta_{s,m}\twonorm{\vct{u}}+\omega(\mathcal{T})+\eta]\Bigg\}\\
\le&\mathbb{P}\Bigg\{\underset{\abs{\mathcal{S}}=s,\text{ }\vct{u}\in\mathcal{T},\vct{v}\in\mathbb{S}^{m-s-1}}{\bigcup}[Y_{(\vct{u},\mathcal{S}),\vct{v}}>\beta_{s,m}\twonorm{\vct{u}}+\omega(\mathcal{T})+\eta]\Bigg\}\\
\le& 2e^{-\frac{\eta^2}{8\sigma^2(\mathcal{T})}}.
\end{align*}
Using the fact that $\twonorm{\mtx{A}_{\mathcal{S}^c}\vct{u}}=\underset{\vct{v}\in\mathbb{S}^{m-s-1}}{\max}\text{ }\vct{v}^*\mtx{I}_{\mathcal{S}^c}\mtx{A}\vct{u}=\underset{\vct{v}\in\mathbb{S}^{m-s-1}}{\max}\text{ }\mtx{X}_{(\vct{u},\mathcal{S}),\vct{v}}$, So

\begin{align*}
\mathbb{P}\Bigg\{\sup_{\mathcal{S}\subset \{1,2,\ldots,m\},\text{ }\abs{\mathcal{S}}= s}\text{ } \twonorm{\mtx{A}_{\mathcal{S}^c}\vct{u}} \ge &\beta_{s,m}\twonorm{\vct{u}}+\omega(\mathcal{T})+\eta\Bigg\}\\=&
\mathbb{P}\Bigg\{\underset{\abs{\mathcal{S}}=s,\text{ }\vct{u}\in\mathcal{T}}{\bigcup}\text{ }\underset{\vct{v}\in\mathbb{S}^{m-s-1}}{\max}\text{ }\mtx{X}_{(\vct{u},\mathcal{S}),\vct{v}}>\twonorm{\vct{u}}\beta_{s,m}+\omega(\mathcal{T})+\eta\Bigg\}\\
=&\mathbb{P}\Bigg\{\underset{\abs{\mathcal{S}}=s,\text{ }\vct{u}\in\mathcal{T},\text{ }\vct{v}\in\mathbb{S}^{m-s-1}}{\bigcup}[X_{(\vct{u},\mathcal{S}),\vct{v}}>\beta_{s,m}\twonorm{\vct{u}}+\omega(\mathcal{T})+\eta]\Bigg\}
\end{align*}
concludes the proof.

Next, we turn our attention to proving \eqref{eq2}. To this aim we begin by stating a lemma due to Gordon \cite{Gor}.
\begin{lemma}\label{Gord}[Gordon's inequality] Let $\left(X_{ij}\right)$ and $\left(Y_{ut}\right)$, $1\le i \le n$, $1\le j\le m$, be Gaussian random vectors. Assume that we have the following inequalities for all choices of indices:
\begin{align}
\label{cond}
\E[X_{i,j}X_{i,k}]\le \E[Y_{i,j}Y_{i,k}]\quad&\text{for all }i,j,k,\nonumber\\
\E[X_{i,j}X_{\ell,k}]\ge \E[Y_{i,j}Y_{\ell,k}]\quad&\text{for all }i\neq \ell\text{ }\text{and}\text{ }j,k,\nonumber\\
\E[X_{i,j}^2]=\E[Y_{i,j}^2]\text{ }\text{ }\quad\quad&\text{for all }i,j.
\end{align}
Then, for all real numbers $\lambda_{i,j}$,
\begin{align*}
\mathbb{P}\bigg\{\underset{i\le n}{\bigcap}\text{ }\underset{j\le m}{\bigcup}[X_{ij}\ge \lambda_{ij}]\bigg\}\ge \mathbb{P}\bigg\{\underset{i\le n}{\bigcap}\text{ }\underset{j\le m}{\bigcup}[Y_{ij}\ge \lambda_{ij}]\bigg\}.
\end{align*}
Consequently,
\begin{align*}
\E\bigg[\underset{i\le n}{\min}\text{ }\underset{j\le m}{\max}\text{ }Y_{ij}\bigg]\le \E\bigg[\underset{i\le n}{\min}\text{ }\underset{j\le m}{\max}\text{ }X_{ij}\bigg].
\end{align*}
\end{lemma}
Define $\mtx{I}_{\mathcal{S}^c}\in\R^{(m-s)\times n}$ as the part of the identity matrix that keeps the rows indexed by $\mathcal{S}^c$. For $\vct{u}\in\mathcal{T}$ and $\vct{v}\in\mathbb{S}^{m-s-1}=\{\vct{v}\in\R^{m-s};\text{ }\twonorm{\vct{v}}=1\}$, we define two Gaussian processes
\begin{align*}
X_{(\vct{u},\mathcal{S}),\vct{v}}=\vct{v}^*\mtx{I}_{\mathcal{S}^c}\mtx{A}\vct{u}+\twonorm{\vct{u}}g\quad\text{and}\quad Y_{(\vct{u},\mathcal{S}),\vct{v}}=\twonorm{\vct{u}}\vct{v}^*\mtx{I}_{\mathcal{S}^c}\vct{a}+\vct{g}^*\vct{u}.%\quad\text{and}\quad Z_{(\vct{u},\mathcal{S}),\vct{v}}=\twonorm{\vct{u}}(\vct{v}^*\mtx{I}_{\mathcal{S}^c}\vct{a}-\alpha_{s,m})+\vct{g}^*\vct{u}.
\end{align*}
Here, $\vct{a}\in\R^m$ is distributed as $\mathcal{N}(\vct{0},\mtx{I}_m)$, $\vct{g}\in\R^n$ is distributed as $\mathcal{N}(\vct{0},\mtx{I}_n)$, and $g\in\R$ is distributed as $\mathcal{N}(0,1)$. The next few steps are essentially identical to the proof of \cite{Gor}[Lemma 3.1] with the text directly borrowed. We mention this part of the argument for the sake of completeness and also to ensure that proper modifications are applied when necessary. Note that
\begin{align*}
\E\big[X_{(\vct{u},\mathcal{S}),\vct{v}}X_{(\tilde{\vct{u}},\tilde{\mathcal{S}}),\tilde{\vct{v}}}\big]-\E\big[Y_{(\vct{u},\mathcal{S}),\vct{v}}Y_{(\tilde{\vct{u}},\tilde{\mathcal{S}}),\tilde{\vct{v}}}\big]=&\langle\vct{u},\tilde{\vct{u}}\rangle\langle\mtx{I}_{\mathcal{S}^c}^T\vct{v},\mtx{I}_{\tilde{\mathcal{S}}^c}^T\tilde{\vct{v}}\rangle+\twonorm{\vct{u}}\twonorm{\tilde{\vct{u}}}-\twonorm{\vct{u}}\twonorm{\tilde{\vct{u}}}\langle\mtx{I}_{\mathcal{S}^c}^T\vct{v},\mtx{I}_{\tilde{\mathcal{S}}^c}^T\tilde{\vct{v}}\rangle\\&-\langle\vct{u},\tilde{\vct{u}}\rangle\\
=&\left(\twonorm{\vct{u}}\twonorm{\tilde{\vct{u}}}-\langle\vct{u},\tilde{\vct{u}}\rangle\right)\left(1-\langle\mtx{I}_{\mathcal{S}^c}^T\vct{v},\mtx{I}_{\tilde{\mathcal{S}}^c}^T\tilde{\vct{v}}\rangle\right)\\
\ge&0
\end{align*}
%It follows that for all $\vct{u},\tilde{\vct{u}}\in\mathcal{T}$, $\vct{v},\tilde{\vct{v}}\in\mathbb{S}^{m-s-1}$ and $\mathcal{S}, \tilde{\mathcal{S}}\subset\{1,2,\ldots,m\}$. 
and equal to zero if $(\vct{u},\mathcal{S})=(\tilde{\vct{u}},\tilde{\mathcal{S}})$ so that the first two inequalities in \eqref{cond} hold. It is also trivial to check that 
\begin{align*}
\E[X_{(\vct{u},\mathcal{S}),\vct{v}}^2]=\E[Y_{(\vct{u},\mathcal{S}),\vct{v}}^2].
\end{align*}
Thus all three inequalities in \eqref{cond} trivially hold. Now note that for each $\vct{u}\in\mathcal{T}$ and $\mathcal{S}\subset\{1,2,\ldots,m\}$ obeying $\abs{\mathcal{S}}=s$ the set
\begin{align*}
\underset{\vct{v}\in\mathbb{S}^{m-s-1}}{\bigcup}[X_{(\vct{u},\mathcal{S}),\vct{v}}\ge \lambda_{\vct{u},\mathcal{S}}]=[\twonorm{\mtx{A}_{\mathcal{S}^c}\vct{u}}+g\twonorm{\vct{u}}\ge \lambda_{\vct{u},\mathcal{S}}]
\end{align*}
is closed in the probability space $\{\R^{mn+1},\mathbb{P}\}$ where $\mathbb{P}$ is the canonical Gaussian measure of $\R^{mn+1}$. Hence 
\begin{align*}
\underset{\abs{\mathcal{S}}=s}{\bigcap}\text{ }\underset{\vct{u}\in\mathcal{T}}{\bigcap}\text{ }\underset{\vct{v}\in\mathbb{S}^{m-s-1}}{\bigcup}\text{ }[X_{(\vct{u},\mathcal{S}),\vct{v}}\ge \lambda_{\vct{u},\mathcal{S}}]
\end{align*}
is closed. The same is true about the corresponding expression with $Y_{(\vct{u},\mathcal{S}),\vct{v}}$. By Lemma \ref{Gord} above, for each finite set $\{(\vct{u}_i,\mathcal{S}_i)\}_1^N\subset\mathcal{T}\times \{1,2,\ldots,m\}$ we have
\begin{align*}
\mathbb{P}\bigg\{\bigcap_{i=1}^N\underset{\vct{v}\in\mathbb{S}^{m-s-1}}{\bigcup}[X_{(\vct{u},\mathcal{S}),\vct{v}}\ge \lambda_{\vct{u},\mathcal{S}}]\bigg\}\ge \mathbb{P}\bigg\{\bigcap_{i=1}^N\underset{\vct{v}\in\mathbb{S}^{m-s-1}}{\bigcup}[Y_{(\vct{u},\mathcal{S}),\vct{v}}\ge \lambda_{\vct{u},\mathcal{S}}]\bigg\}
\end{align*}
and so, ordering the collection of finite subsets of $\mathcal{T}\times \{1,2,\ldots,m\}$ (denoted by $\mathcal{F}$) by inclusion, we obtain that the limits exist and satisfy the inequality
\begin{align*}
\underset{\mathcal{F}}{\lim}\text{ }\mathbb{P}\bigg\{\bigcap_{i=1}^N\underset{\vct{v}\in\mathbb{S}^{m-s-1}}{\bigcup}[X_{(\vct{u},\mathcal{S}),\vct{v}}\ge \lambda_{\vct{u},\mathcal{S}}]\bigg\}\ge \underset{\mathcal{F}}{\lim}\text{ }\mathbb{P}\bigg\{\bigcap_{i=1}^N\underset{\vct{v}\in\mathbb{S}^{m-s-1}}{\bigcup}[Y_{(\vct{u},\mathcal{S}),\vct{v}}\ge \lambda_{\vct{u},\mathcal{S}}]\bigg\}.
\end{align*}
Now using the fact that the sets 
\begin{align*}
\underset{\abs{\mathcal{S}}=s}{\bigcap}\text{ }\underset{\vct{u}\in\mathcal{T}}{\bigcap}\text{ }\underset{\vct{v}\in\mathbb{S}^{m-s-1}}{\bigcup}\text{ }[X_{(\vct{u},\mathcal{S}),\vct{v}}\ge \lambda_{\vct{u},\mathcal{S}}]\quad\text{and}\quad 
\underset{\abs{\mathcal{S}}=s}{\bigcap}\text{ }\underset{\vct{u}\in\mathcal{T}}{\bigcap}\text{ }\underset{\vct{v}\in\mathbb{S}^{m-s-1}}{\bigcup}\text{ }[Y_{(\vct{u},\mathcal{S}),\vct{v}}\ge \lambda_{\vct{u},\mathcal{S}}]
\end{align*}
are closed and $\mathbb{P}$ is a regular measure, it follows easily that the two respective limits over $\mathcal{F}$ are equal to and satisfy the inequality
\begin{align*}
\mathbb{P}\bigg\{\underset{\abs{\mathcal{S}}=s}{\bigcap}\text{ }\underset{\vct{u}\in\mathcal{T}}{\bigcap}\text{ }\underset{\vct{v}\in\mathbb{S}^{m-s-1}}{\bigcup}\text{ }[X_{(\vct{u},\mathcal{S}),\vct{v}}\ge \lambda_{\vct{u},\mathcal{S}}]\bigg\}\ge\mathbb{P}\bigg\{\underset{\abs{\mathcal{S}}=s}{\bigcap}\text{ }\underset{\vct{u}\in\mathcal{T}}{\bigcap}\text{ }\underset{\vct{v}\in\mathbb{S}^{m-s-1}}{\bigcup}\text{ }[Y_{(\vct{u},\mathcal{S}),\vct{v}}\ge \lambda_{\vct{u},\mathcal{S}}]\bigg\}.%\footnote{This concludes the part that is borrowed from \cite{Gord}.}
\end{align*}
This immediately implies that
\begin{align*}
%\label{probtemp}
\mathbb{P}\bigg\{\underset{\abs{\mathcal{S}}=s}{\bigcap}\text{ }\underset{\vct{u}\in\mathcal{T}}{\bigcap}\text{ }[\twonorm{\mtx{A}_{\mathcal{S}^c}\vct{u}}+g\twonorm{\vct{u}}\ge \lambda_{\vct{u},\mathcal{S}}]\bigg\}\ge\mathbb{P}\bigg\{\underset{\abs{\mathcal{S}}=s}{\bigcap}\text{ }\underset{\vct{u}\in\mathcal{T}}{\bigcap}\text{ }[\twonorm{\vct{u}}\twonorm{\vct{a}_{\mathcal{S}^c}}+\vct{g}^*\vct{u}\ge \lambda_{\vct{u},\mathcal{S}}]\bigg\}.
\end{align*}
Now setting
\begin{align*}
\lambda_{\vct{u},\mathcal{S}}=\alpha_{s,m}\twonorm{\vct{u}}-\left(\omega(\mathcal{T})+\eta\right),
\end{align*}
we conclude that 
\begin{align}
\label{probtemp2}
\mathbb{P}\bigg\{\underset{\abs{\mathcal{S}}=s}{\bigcap}\text{ }\underset{\vct{u}\in\mathcal{T}}{\bigcap}\text{ }[\twonorm{\mtx{A}_{\mathcal{S}^c}\vct{u}}+g\twonorm{\vct{u}}&\ge \alpha_{s,m}\twonorm{\vct{u}}-\left(\omega(\mathcal{T})+\eta\right)]\bigg\}\nonumber\\
&\ge\mathbb{P}\bigg\{\underset{\abs{\mathcal{S}}=s}{\bigcap}\text{ }\underset{\vct{u}\in\mathcal{T}}{\bigcap}\text{ }[\twonorm{\vct{u}}\twonorm{\vct{a}_{\mathcal{S}^c}}+\vct{g}^*\vct{u}\ge \alpha_{s,m}\twonorm{\vct{u}}-\left(\omega(\mathcal{T})+\eta\right)]\bigg\}\nonumber\\
&=\mathbb{P}\bigg\{\underset{\vct{u}\in\mathcal{T}}{\bigcap}\text{ }[\twonorm{\vct{u}}\underset{\abs{\mathcal{S}}=s}{\inf}\twonorm{\vct{a}_{\mathcal{S}^c}}+\vct{g}^*\vct{u}\ge \alpha_{s,m}\twonorm{\vct{u}}-\left(\omega(\mathcal{T})+\eta\right)]\bigg\}.
\end{align}

Since taking infimum over a set is equivalent to taking intersection over all elements of that set, we can write

\begin{align}
\label{eqq1}
\mathbb{P}\Bigg\{\underset{\vct{u}\in\mathcal{T}}{\bigcap}\text{ }\bigg[\vct{g}^*\vct{u}\ge -\left(\omega(\mathcal{T})+\frac{\eta}{2}\right)\bigg]\Bigg\}=&\mathbb{P}\bigg\{\underset{\vct{u}\in\mathcal{T}}{\inf}\text{ }\vct{g}^*\vct{u}\ge -\left(\omega(\mathcal{T})+\frac{\eta}{2}\right)\bigg\}\nonumber\\
=&\mathbb{P}\bigg\{-\underset{\vct{u}\in\mathcal{T}}{\sup}\text{ }-\vct{g}^*\vct{u}\ge -\left(\omega(\mathcal{T})+\frac{\eta}{2}\right)\bigg\},\nonumber\\
=&\mathbb{P}\bigg\{\underset{\vct{u}\in\mathcal{T}}{\sup}\text{ }-\vct{g}^*\vct{u}\le \left(\omega(\mathcal{T})+\frac{\eta}{2}\right)\bigg\},\nonumber\\
\ge&1-e^{-\frac{\eta^2}{8\sigma^2(\mathcal{T})}}.
\end{align}

In the last inequality, we used the concentration of Lipschitz functions of Gaussians and the definition of Gaussian width.

Now define the function $g(\vct{x})=\underset{\mathcal{S}\subset \{1,2,\ldots,m\},\text{ }\abs{\mathcal{S}}=s}{\inf}\text{ }\twonorm{\vct{x}_{\mathcal{S}^c}}$ and let 

\begin{align*}
\mathcal{S}_{\vct{x}}^c=\underset{\mathcal{S}\subset \{1,2,\ldots,m\},\text{ }\abs{\mathcal{S}}=s}{\arg\min} \twonorm{\vct{x}_{\mathcal{S}^c}}\quad\text{and}\quad\mathcal{S}_{\vct{y}}^c=&\underset{\mathcal{S}\subset \{1,2,\ldots,m\},\text{ }\abs{\mathcal{S}}=s}{\arg\min} \twonorm{\vct{y}_{\mathcal{S}^c}}.
\end{align*}

We claim that $g(\vct{x})$ is a Lipschitz function and then we can utilize the concentration of measure for Gaussian random variables. Without loss of generality we assume $g(\vct{x})\ge g(\vct{y})$. Thus,

\begin{align*}
\abs{g(\vct{x})-g(\vct{y})}=g(\vct{x})-g(\vct{y})=\twonorm{\vct{x}_{\mathcal{S}_{\vct{x}}^c}}-\twonorm{\vct{y}_{\mathcal{S}_{\vct{y}}^c}}\le \twonorm{\vct{x}_{\mathcal{S}_{\vct{y}}^c}}-\twonorm{\vct{y}_{\mathcal{S}_{\vct{y}}^c}}\le \twonorm{\left(\vct{x}-\vct{y}\right)_{\mathcal{S}_{\vct{y}}^c}}\le\twonorm{\vct{x}-\vct{y}}.
\end{align*}

Hence $g(\vct{a})$ is a Lipschitz function of a Gaussian random variable. Thus, the random variable $Z:=g(\vct{a})$ obeys
\begin{align*}
\text{Var}(Z)\le 1\quad\Rightarrow\quad \E[Z^2]-\left(\E[Z]\right)^2\le1\quad\Rightarrow\quad \E[Z]\ge\sqrt{\E[Z^2]-1}.
\end{align*}

Therefore, 

\begin{align}\label{infa}
\E\bigg[\underset{\mathcal{S}\subset \{1,2,\ldots,m\},\text{ }\abs{\mathcal{S}}=s}{\inf}\text{ }\twonorm{\vct{a}_{\mathcal{S}^c}}\bigg]\ge&\sqrt{\E\bigg[\underset{\mathcal{S}\subset \{1,2,\ldots,m\},\text{ }\abs{\mathcal{S}}=s}{\inf}\text{ }\twonorm{\vct{a}_{\mathcal{S}^c}}^2\bigg]-1}\nonumber\\
\ge&\sqrt{\E\bigg[\twonorm{\vct{a}}^2-\underset{\mathcal{S}\subset \{1,2,\ldots,m\},\text{ }\abs{\mathcal{S}}=s}{\sup}\text{ }\twonorm{\vct{a}_{\mathcal{S}}}^2\bigg]-1}\nonumber\\
=&\sqrt{m-1-\E\bigg[\underset{\mathcal{S}\subset \{1,2,\ldots,m\},\text{ }\abs{\mathcal{S}}=s}{\sup}\text{ }\twonorm{\vct{a}_{\mathcal{S}}}^2\bigg]}\nonumber\\
\ge&\sqrt{m-2-\E\bigg[\underset{\mathcal{S}\subset \{1,2,\ldots,m\},\text{ }\abs{\mathcal{S}}=s}{\sup}\text{ }\twonorm{\vct{a}_{\mathcal{S}}}\bigg]^2}
\end{align}

In the last line we used the fact that $\underset{\mathcal{S}\subset \{1,2,\ldots,m\},\text{ }\abs{\mathcal{S}}=s}{\sup}\text{ }\twonorm{\vct{a}_{\mathcal{S}}}$ is a Lipschitz function of $\vct{a}$ and therefore the random variable $X:=\underset{\mathcal{S}\subset \{1,2,\ldots,m\},\text{ }\abs{\mathcal{S}}=s}{\sup}\text{ }\twonorm{\vct{a}_{\mathcal{S}}}$ obeys
\begin{align*}
\text{Var}(X)\le 1\quad\Rightarrow\quad \E[X^2]\le\left(\E[X]\right)^2+1.
\end{align*}
We now wish to prove that

\begin{align}\label{expbound}
\E\bigg[\underset{\mathcal{S}\subset \{1,2,\ldots,m\},\text{ }\abs{\mathcal{S}}=s}{\sup}\text{ }\twonorm{\vct{a}_{\mathcal{S}}}\bigg]\le\sqrt{5s\log\left({\frac{em}{s}}\right)}.
\end{align}

To this aim first note that using \eqref{mmmtemp} with changing $m-s$ to $s$ and $\mathcal{S}^c$ to $\mathcal{S}$ we have

\begin{align*}
\mathbb{P}\Bigg\{\underset{\mathcal{S}\subset \{1,2,\ldots,m\},\text{ }\abs{\mathcal{S}}=s}{\sup}\text{ }\twonorm{\vct{a}_{\mathcal{S}}}\ge\eta+\sqrt{s}+\sqrt{2s\log\left(\frac{em}{s}\right)}\Bigg\}
\le e^{\frac{-\eta^2}{2}}.
\end{align*}
To bound the expected value we use the tail bound above together with the fact that $s\ge 1$ ($s=0$ is trivial) to conclude that
\begin{align*}\label{expbound1}
\E[X]=&\int_{0}^{+\infty} \mathbb{P}\{X>t\} dt\nonumber\\=&\int_{0}^{\sqrt{s}+\sqrt{2s\log\left(\frac{em}{s}\right)}} \mathbb{P}\{X>t\} dt+\int_{\sqrt{s}+\sqrt{2(s)\log\left(\frac{em}{s}\right)}}^{+\infty} \mathbb{P}\{X>t\} dt\nonumber\\\le&\sqrt{s}+\sqrt{2s\log\left(\frac{em}{s}\right)}+\int_{\sqrt{s}+\sqrt{2s\log\left(\frac{em}{s}\right)}}^{+\infty}  e^{\frac{-\eta^2}{2}}dt\nonumber\\\le&\sqrt{s}+\sqrt{2s\log\left(\frac{em}{s}\right)}+\int_{0}^{+\infty}  e^{\frac{-\eta^2}{2}}dt\nonumber\\\le&\sqrt{s}+\sqrt{2s\log\left(\frac{em}{s}\right)}+\sqrt{\frac{\pi}{2}}\nonumber\\
\le&\sqrt{s}+\sqrt{2s\log\left(\frac{em}{s}\right)}+\sqrt{\frac{\pi s}{2}}\nonumber\\\le&\sqrt{5s\log\left({\frac{em}{s}}\right)}.
\end{align*}
concluding the proof of \eqref{expbound}.

Combining \eqref{infa} with \eqref{expbound} we arrive at

\begin{align}
\E\bigg[\underset{\mathcal{S}\subset \{1,2,\ldots,m\},\text{ }\abs{\mathcal{S}}=s}{\inf}\text{ }\twonorm{\vct{a}_{\mathcal{S}^c}}\bigg]\ge&\sqrt{m-2-5s\log(\frac{em}{s})}:=\alpha_{s,m}
\end{align}

%by substituting $t=\eta+\sqrt{s}+\sqrt{2(s)\log\left(\frac{em}{s}\right)}$

%\begin{align}\label{expbound}
%\E[X]=&\int_{-(\sqrt{s}+\sqrt{2(s)\log\left(\frac{em}{s}\right)})}^{+\infty} \mathbb{P}(X>\eta+\sqrt{s}+\sqrt{2(s)\log\left(\frac{em}{s}\right)}) d\eta\\\le&\int_{-(\sqrt{s}+\sqrt{2(s)\log\left(\frac{em}{s}\right)})}^{+\infty} e^{\frac{-\eta^2}{2}} dt\\=&\sqrt{\frac{\pi}{2}}+\sqrt{\frac{\pi}{2}}\text{erf}(\frac{\sqrt{s}+\sqrt{2s\log(\frac{em}{s})}}{\sqrt{2}})
%\end{align}

%&\sqrt{m-1-4s\log\left(\frac{em}{s}\right)}\nonumber\\
%=&\alpha_{s,m}.

As mentioned earlier, $g(\vct{a})$ is Lipschitz function of $\vct{a}$. Thus, by concentration of Lipschitz functions of Gaussians we have
\begin{align}\label{conc}
\mathbb{P}\bigg\{g(\vct{a})\ge \E[g(\vct{a})]-\frac{\eta}{2\sigma(\mathcal{T})}\bigg\}\ge 1-e^{-\frac{\eta^2}{8\sigma^2(\mathcal{T})}},
\end{align}

Using the fact that $\frac{\twonorm{\vct{u}}}{\sigma(\mathcal{T})}\le1$ and together with \eqref{infa} as well as \eqref{conc}, we can deduce that

\begin{align}
\label{eqq2}
\mathbb{P}\Bigg\{\underset{\vct{u}\in\mathcal{T}}{\bigcap}\text{ }\bigg[\twonorm{\vct{u}}\inf_{\abs{\mathcal{S}}=s}\twonorm{\vct{a}_{\mathcal{S}^c}}\ge \twonorm{\vct{u}}&\alpha_{s,m}-\frac{\eta}{2}\bigg]\Bigg\}\\\ge&
\mathbb{P}\Bigg\{\underset{\vct{u}\in\mathcal{T}}{\bigcap}\text{ }\bigg[\twonorm{\vct{u}}\inf_{\abs{\mathcal{S}}=s}\twonorm{\vct{a}_{\mathcal{S}^c}}\ge \twonorm{\vct{u}}\E\bigg[\inf_{\abs{\mathcal{S}}=s}\twonorm{\vct{a}_{\mathcal{S}^c}}\bigg]-\frac{\eta}{2}\bigg]\Bigg\}\nonumber\\
\ge&\mathbb{P}\Bigg\{\underset{\vct{u}\in\mathcal{T}}{\bigcap}\text{ }\bigg[\twonorm{\vct{u}}g(\vct{a})\ge \twonorm{\vct{u}}\E[g(\vct{a})]-\twonorm{\vct{u}}\frac{\eta}{2\sigma(\mathcal{T})}\bigg]\Bigg\}\nonumber\\
\ge& 1-e^{-\frac{\eta^2}{8\sigma^2(\mathcal{T})}}.
\end{align}

So far we have obtained lower bounds on the probability of sets $\Bigg\{\underset{\vct{u}\in\mathcal{T}}{\bigcap}\text{ }\bigg[\vct{g}^*\vct{u}\ge -\left(\omega(\mathcal{T})+\frac{\eta}{2}\right)\bigg]\Bigg\}$ and $\Bigg\{\underset{\vct{u}\in\mathcal{T}}{\bigcap}\text{ }\bigg[\twonorm{\vct{u}}\inf_{\abs{\mathcal{S}}=s}\twonorm{\vct{a}_{\mathcal{S}^c}}\ge \twonorm{\vct{u}}\alpha_{s,m}-\frac{\eta}{2}\bigg]\Bigg\}$
. In the following we aim to employ these two lower bounds. Note that if

\begin{align*}
\underset{\vct{u}\in\mathcal{T}}{\inf}\text{ }\left(\twonorm{\vct{u}}\left(\underset{\abs{\mathcal{S}}=s}{\inf}\twonorm{\vct{a}_{\mathcal{S}^c}}-\alpha_{s,m}\right)+\vct{g}^*\vct{u}\right)<-\left(\omega(\mathcal{T})+\eta\right),
\end{align*}

then we have either $\quad \underset{\vct{u}\in\mathcal{T}}{\inf}\text{ }\twonorm{\vct{u}}\left(\underset{\abs{\mathcal{S}}=s}{\inf}\twonorm{\vct{a}_{\mathcal{S}^c}}-\alpha_{s,m}\right)<-\frac{\eta}{2}\quad $ or $\quad \underset{\vct{u}\in\mathcal{T}}{\inf} \text{ }\left(\vct{g}^*\vct{u}\right)<-\left(\omega(\mathcal{T})+\frac{\eta}{2}\right).$
%\begin{align*}
%\underset{\vct{u}\in\mathcal{T}}{\inf}\text{ }\twonorm{\vct{u}}\left(\underset{\abs{\mathcal{S}}=s}{\inf}\twonorm{\vct{a}_{\mathcal{S}^c}}-\alpha_{s,m}\right)<-\frac{\eta}{2},
%\end{align*}
%or
%\begin{align*}
%\underset{\vct{u}\in\mathcal{T}}{\inf} \text{ }\left(\vct{g}^*\vct{u}\right)<-\left(\omega(\mathcal{T})+\frac{\eta}{2}\right).
%\end{align*}
This implies that

\begin{align*}
\bigg\{\vct{a},\vct{g}:\text{ }\underset{\vct{u}\in\mathcal{T}}{\inf}\text{ }\left(\twonorm{\vct{u}}\left(\underset{\abs{\mathcal{S}}=s}{\inf}\twonorm{\vct{a}_{\mathcal{S}^c}}-\alpha_{s,m}\right)+\vct{g}^*\vct{u}\right)<-\left(\omega(\mathcal{T})+\eta\right)\bigg\},
\end{align*}

is a subset of 

\begin{align*}
\bigg\{\vct{a},\vct{g}:\text{ }\underset{\vct{u}\in\mathcal{T}}{\inf}\text{ }\twonorm{\vct{u}}\left(\underset{\abs{\mathcal{S}}=s}{\inf}\twonorm{\vct{a}_{\mathcal{S}^c}}-\alpha_{s,m}\right)<-\frac{\eta}{2}\bigg\}\bigcup\bigg\{\vct{a},\vct{g}:\text{ }\underset{\vct{u}\in\mathcal{T}}{\inf} \text{ }\left(\vct{g}^*\vct{u}\right)<-\left(\omega(\mathcal{T})+\frac{\eta}{2}\right)\bigg\}.
\end{align*}

The latter is equivalent to

\begin{align*}
\underset{\vct{u}\in\mathcal{T}}{\bigcup}\big[\twonorm{\vct{u}}\underset{\abs{\mathcal{S}}=s}{\inf}&\twonorm{\vct{a}_{\mathcal{S}^c}}+\vct{g}^*\vct{u}< \alpha_{s,m}\twonorm{\vct{u}}-\left(\omega(\mathcal{T})+\eta\right)\big]\\
%&\subset \underset{\vct{u}\in\mathcal{T}}{\bigcap}\Bigg(\bigg[\twonorm{\vct{u}}\inf_{\abs{\mathcal{S}}=s}\twonorm{\vct{a}_{\mathcal{S}^c}}< \twonorm{\vct{u}}\alpha_{s,m}-\frac{\eta}{2}\bigg]\bigcup \bigg[\vct{g}^*\vct{u}< -\left(\omega(\mathcal{T})+\frac{\eta}{2}\right)\bigg]\Bigg),\\
&\subset\left(\underset{\vct{u}\in\mathcal{T}}{\bigcup}\bigg[\twonorm{\vct{u}}\inf_{\abs{\mathcal{S}}=s}\twonorm{\vct{a}_{\mathcal{S}^c}}< \twonorm{\vct{u}}\alpha_{s,m}-\frac{\eta}{2}\bigg]\right)\bigcup\left(\underset{\vct{u}\in\mathcal{T}}{\bigcup}\bigg[\vct{g}^*\vct{u}< -\left(\omega(\mathcal{T})+\frac{\eta}{2}\right)\bigg]\right).
\end{align*}

Considering the probability of these sets, we have

\begin{align*}
\mathbb{P}\bigg\{\underset{\vct{u}\in\mathcal{T}}{\bigcup}\big[\twonorm{\vct{u}}\underset{\abs{\mathcal{S}}=s}{\inf}&\twonorm{\vct{a}_{\mathcal{S}^c}}+\vct{g}^*\vct{u}< \alpha_{s,m}\twonorm{\vct{u}}-\left(\omega(\mathcal{T})+\eta\right)\big]\bigg\}\\
&\le\mathbb{P}\bigg\{\underset{\vct{u}\in\mathcal{T}}{\bigcup}\bigg[\twonorm{\vct{u}}\inf_{\abs{\mathcal{S}}=s}\twonorm{\vct{a}_{\mathcal{S}^c}}< \twonorm{\vct{u}}\alpha_{s,m}-\frac{\eta}{2}\bigg]\bigg\}+\mathbb{P}\bigg\{\underset{\vct{u}\in\mathcal{T}}{\bigcup}\bigg[\vct{g}^*\vct{u}< -\left(\omega(\mathcal{T})+\frac{\eta}{2}\right)\bigg]\bigg\}.
\end{align*}

Taking complements of both sides and using the bounds from \eqref{eqq1} and \eqref{eqq2} we conclude that

\begin{align}
\mathbb{P}\bigg\{\underset{\vct{u}\in\mathcal{T}}{\bigcap}\big[\twonorm{\vct{u}}\underset{\abs{\mathcal{S}}=s}{\inf}&\twonorm{\vct{a}_{\mathcal{S}^c}}+\vct{g}^*\vct{u}\ge \alpha_{s,m}\twonorm{\vct{u}}-\left(\omega(\mathcal{T})+\eta\right)\big]\bigg\}\nonumber\\
&\ge\mathbb{P}\bigg\{\underset{\vct{u}\in\mathcal{T}}{\bigcap}\bigg[\twonorm{\vct{u}}\inf_{\abs{\mathcal{S}}=s}\twonorm{\vct{a}_{\mathcal{S}^c}}\ge \twonorm{\vct{u}}\alpha_{s,m}-\frac{\eta}{2}\bigg]\bigg\}+\mathbb{P}\bigg\{\underset{\vct{u}\in\mathcal{T}}{\bigcap}\bigg[\vct{g}^*\vct{u}\ge -\left(\omega(\mathcal{T})+\frac{\eta}{2}\right)\bigg]\bigg\}-1\nonumber\\
&\ge 1-2e^{-\frac{\eta^2}{8\sigma^2(\mathcal{T})}}.
\end{align}

The latter inequality together with \eqref{probtemp2} implies that
\begin{align}
\label{fintemp}
\mathbb{P}\bigg\{\underset{\abs{\mathcal{S}}=s}{\inf}\text{ }\underset{\vct{u}\in\mathcal{T}}{\inf}\text{ }(\twonorm{\mtx{A}_{\mathcal{S}^c}\vct{u}}+g\twonorm{\vct{u}}-&\alpha_{s,m}\twonorm{\vct{u}})\ge -\left(\omega(\mathcal{T})+\eta\right)\bigg\}\nonumber\\
=&\mathbb{P}\bigg\{\underset{\abs{\mathcal{S}}=s}{\bigcap}\text{ }\underset{\vct{u}\in\mathcal{T}}{\bigcap}\text{ }[\twonorm{\mtx{A}_{\mathcal{S}^c}\vct{u}}+g\twonorm{\vct{u}}-\alpha_{s,m}\twonorm{\vct{u}}\ge -\left(\omega(\mathcal{T})+\eta\right)]\bigg\}\nonumber\\
\ge& 1-2e^{-\frac{\eta^2}{8\sigma^2(\mathcal{T})}}.
\end{align}

In order to find the relationship between the probability of the latter set with the probability of the set defined in \eqref{eq2}, we define the following three probabilities 
\begin{align*}
p=&\mathbb{P}\bigg\{\underset{\abs{\mathcal{S}}=s}{\inf}\text{ }\underset{\vct{u}\in\mathcal{T}}{\inf}\text{ }\left(\twonorm{\mtx{A}_{\mathcal{S}^c}\vct{u}}+g\twonorm{\vct{u}}-\alpha_{s,m}\twonorm{\vct{u}}\right)\ge -\left(\omega(\mathcal{T})+\eta\right)\bigg\},\\
p_{-}=&\mathbb{P}\bigg\{\underset{\abs{\mathcal{S}}=s}{\inf}\text{ }\underset{\vct{u}\in\mathcal{T}}{\inf}\text{ }\left(\twonorm{\mtx{A}_{\mathcal{S}^c}\vct{u}}+g\twonorm{\vct{u}}-\alpha_{s,m}\twonorm{\vct{u}}\right)\ge -\left(\omega(\mathcal{T})+\eta\right)\text{ }\bigg|\text{ } g\le 0\bigg\},\\
p_{+}=&\mathbb{P}\bigg\{\underset{\abs{\mathcal{S}}=s}{\inf}\text{ }\underset{\vct{u}\in\mathcal{T}}{\inf}\text{ }\left(\twonorm{\mtx{A}_{\mathcal{S}^c}\vct{u}}+g\twonorm{\vct{u}}-\alpha_{s,m}\twonorm{\vct{u}}\right)\ge -\left(\omega(\mathcal{T})+\eta\right)\text{ }\bigg|\text{ } g> 0\bigg\},\\
p_0=&\mathbb{P}\bigg\{\underset{\abs{\mathcal{S}}=s}{\inf}\text{ }\underset{\vct{u}\in\mathcal{T}}{\inf}\text{ }\left(\twonorm{\mtx{A}_{\mathcal{S}^c}\vct{u}}-\alpha_{s,m}\twonorm{\vct{u}}\right)\ge -\left(\omega(\mathcal{T})+\eta\right)\text{ }\bigg\}.
\end{align*}
Now note that by the above definitions and the independence of $\mtx{A}$ and $g$ we can conclude that
\begin{align}
\label{tp}
1\ge p_{+}\ge p_0\ge p_{-}.
\end{align}
By the law of total probability $p=\frac{p_{-}+p_{+}}{2}$. Now using the fact that $p_{+}\le 1$ together with \eqref{tp} we can conclude that
\begin{align*}
1-p=\frac{1-p_{-}}{2}+\frac{1-p_{+}}{2}\ge \frac{1-p_{-}}{2}\ge \frac{1-p_0}{2}\quad\Rightarrow\quad p_0\ge 2p-1.
\end{align*}
The latter inequality together with \eqref{fintemp} implies that
\begin{align*}
\mathbb{P}\bigg\{\underset{\abs{\mathcal{S}}=s}{\inf}\text{ }\underset{\vct{u}\in\mathcal{T}}{\inf}\text{ }\left(\twonorm{\mtx{A}_{\mathcal{S}^c}\vct{u}}+g\twonorm{\vct{u}}-\alpha_{s,m}\twonorm{\vct{u}}\right)\ge -\left(\omega(\mathcal{T})+\eta\right)\bigg\}\ge 1-4e^{-\frac{\eta^2}{8\sigma^2(\mathcal{T})}},
\end{align*}
concluding the proof of \eqref{eq2}.
% show that $\twonorm{\mtx{A}\vct{u}}\ge b_m\twonorm{\vct{u}}-\omega(\mathcal{T})-\eta$. To accomplish this, we make use of Lemma $5.1$ of \cite{OymLAS}. The following is an immediate corollary.
%%\begin{corollary} Let $\mtx{A}\in\R^{m\times n},\vct{g}\in\R^n,\vct{h}\in\R^m$ be independent vectors with independent $\mathcal{N}(0,1)$ entries. Then, for any $c\in\R$
%%\begin{align}
%%\mathbb{P}(\min_{\vct{u}\in \mathcal{T}}\max_{\vct{v}\in\mathbb{S}^{m-1}}\vct{v}^*\mtx{A}\vct{u}-b_m\twonorm{\vct{u}}\geq c)\geq 2\mathbb{P}(\min_{\vct{u}\in\mathcal{T}}\max_{\vct{v}\in\mathbb{S}^{m-1}}\vct{v}^*\vct{h}\twonorm{\vct{u}}-\vct{u}^*\vct{g}\twonorm{\vct{v}}-b_m\twonorm{\vct{u}}\geq c).\label{two times}
%%\end{align}
%%\end{corollary}
%%The right-hand side can be simplified to $\min_{\vct{u}\in\mathcal{T}}(\twonorm{\vct{h}}-b_m)\twonorm{\vct{u}}-\vct{u}^*\vct{g}$, which is $\sqrt{2}\sigma(\mathcal{T})$-Lipschitz function of the vector $\begin{bmatrix}\vct{g}^*&\vct{h}^*\end{bmatrix}$. As a result we have
%%\begin{align}
%%\mathbb{P}(\min_{\vct{u}\in\mathcal{T}}(\twonorm{\vct{h}}-b_m)\twonorm{\vct{u}}-\vct{u}^*\vct{g}\geq -\omega(\mathcal{T})-\eta)\geq 1-\exp(-\frac{\eta^2}{4\sigma^2(\mathcal{T})}).
%%\end{align}
%%The proof is complete by combining the latter with \eqref{two times}.

%\bibliography{Bibfiles-2}
%\bibliographystyle{icml2018}

\end{document}